\documentclass[twocolumn]{aastex631}

\shorttitle{Isolated BHs from single and binary origin}
\shortauthors{Vigna-G\'omez \& Ramirez-Ruiz}
\graphicspath{{./}{plots/}{appendixPlots/}}

\usepackage[fleqn]{amsmath} 
\usepackage{amssymb}  
\usepackage{gensymb}
\usepackage[nolist]{acronym}
\usepackage{ulem}
\usepackage{bm}
\usepackage{etoolbox}
\usepackage{hyperref}
\usepackage{enumitem}

\definecolor{ochre}{rgb}{0.8, 0.47, 0.13}

\newcommand\event{OB110462}
\newcommand\confidence{95\%\ }
\newcommand\mass{10}

\usepackage{import}
\usepackage{xspace}
\newcommand{\monei}{\ensuremath{m_{1,\rm{i}}}\xspace}
\newcommand{\mtwoi}{\ensuremath{m_{2,\rm{i}}}\xspace}

\newcommand{\ai}{\ensuremath{a_{\rm{i}}}\xspace}
\newcommand{\qi}{\ensuremath{q_{\rm{i}}}\xspace}
\newcommand{\Zi}{\ensuremath{Z_{\rm{i}}}\xspace}
\newcommand{\vk}{\ensuremath{v_{\rm{k}}}\xspace}
\newcommand{\thetak}{\ensuremath{{\theta}_{\rm{k}}}\xspace}

\newcommand{\ei}{\ensuremath{{e}_{\rm{i}}}\xspace}

\newcommand{\Msun}{\ensuremath{\,\rm{M}_{\odot}}\xspace}

\newcommand{\kms}{\ensuremath{\,\rm{km}\,\rm{s}^{-1}}\xspace}
\newcommand{\AU}{\ensuremath{\,\mathrm{AU}}\xspace}

\begin{document}

\title{A binary origin for the first isolated stellar-mass black hole detected with astrometric microlensing}

\correspondingauthor{Alejandro Vigna-G\'omez}
\email{avigna@mpa-garching.mpg.de}

\author[0000-0003-1817-3586]{Alejandro Vigna-G\'omez}
\affiliation{Max-Planck-Institut f\"ur Astrophysik, Karl-Schwarzschild-Str. 1, D-85748 Garching, Germany}
\affiliation{Niels Bohr International Academy, Niels Bohr Institute, University of Copenhagen, Blegdamsvej 17, DK-2100,  Copenhagen, Denmark}

\author[0000-0003-2558-3102]{Enrico Ramirez-Ruiz}
\affiliation{Department of Astronomy and Astrophysics,
University of California,
Santa Cruz, CA 95064, USA}



\begin{abstract}

The Milky Way is believed to host hundreds of millions of quiescent stellar-mass black holes (BHs). In the last decade, some of these objects have been potentially uncovered via gravitational microlensing events.
All these detections resulted in a degeneracy between the velocity and the mass of the lens. 
This degeneracy has been lifted, for the first time, with the recent astrometric microlensing detection of \event.
However, two independent studies reported very different lens mass for this event.
\cite{2022ApJ...933...83S} inferred a lens mass of $7.1 \pm 1.3\ \rm{M_\odot}$, consistent with a BH, while \cite{2022ApJ...933L..23L} inferred $1.6 - 4.2\ \rm{M_\odot}$, consistent with either a neutron star or a BH.
Here, we study the landscape of isolated BHs formed in the field. 
In particular, we focus on the mass and center-of-mass speed of four sub-populations: isolated BHs from single-star origin, disrupted BHs of binary-star origin, main-sequence stars with a compact object companion, and double compact object mergers.
Our model predicts that most ($\gtrsim 70\%$) isolated BHs in the Milky Way are of binary origin.
However, non-interactions lead to most massive BHs ($\gtrsim 15-20\ \rm{M_\odot}$) being predominantly of single origin.
Under the assumption that \event\ is a free-floating compact object we conclude that it is more likely to be a BH originally belonging to a binary system.
Our results suggest that low-mass BH microlensing events can be useful to understand binary evolution of massive stars in the Milky Way, while high-mass BH lenses can be useful to probe single stellar evolution.
\end{abstract}



\section{Introduction} \label{sec:intro}

Isolated stellar-mass black holes (BHs) are expected to be abundant, with $\approx 10^8$ initially estimated to reside in the Milky Way \citep{1983bhwd.book.....S}. Yet their detection remains elusive.
For decades, gravitational microlensing  has been the most promising avenue to detecting quiescent stellar-mass compact-objects in the Milky Way \citep[e.g.,][]{1986ApJ...304....1P,1996ARA&A..34..419P}, but because  the alignment needed is so accurate and tough to  foretell, microlensing is highly infrequent. 
Given that $\approx 10^8$ BHs are predicted to be drifting through the Milky Way, the probability of seeing a transient brightening of a background star when an isolated BH flits across the field of view \citep[e.g.,][]{2016MNRAS.458.3012W,2020A&A...636A..20W} could be high in wide and deep sky surveys.

Several surveys now routinely search for these microlensing events, including the Optical Gravitational Lensing Experiment (OGLE) and the Microlensing Observations in Astrophysics (MOA) survey. These two surveys independently  reported \citep{2022ApJ...933...83S, 2022ApJ...933L..23L} the detection of a compact object using, for the first time, astrometric microlensing of target OGLE-2011-BLG-0462/MOA-2011-BLG-191 (\event). However, the reported properties, and particularly the lens mass, were significantly different in each study.
\cite{2022ApJ...933...83S} reported \event\ as a BH detection with lens mass of $7.1 \pm 1.3\ \rm{M_\odot}$, at a distance of $1.58 \pm 0.18$ kpc, and with transverse space velocity of $\approx 45\ \rm{km\ s^{-1}}$.
Additionally, \cite{2022ApJ...933...83S}  ruled out a stellar or compact object companion within $\approx 0.18-230$ AU.
Alternatively, \cite{2022ApJ...933L..23L} reported \event\ as either a neutron star (NS) or a BH with lens mass between $1.6 - 4.2\ \rm{M_\odot}$, at a distance between  $0.69 - 1.75$  kpc, and with transverse motion $< 25\ \rm{km\ s^{-1}}$.
While both studies reported \event\ as an isolated compact object in the direction of the Galactic bulge, the apparent free-floating nature might not be representative of their origin.

Here, we explore the landscape of compact objects in the context of microlensing targets using binary population synthesis models framed to explain the number density of compact binary mergers as inferred by ground-based gravitational-wave observatories. 
We focus on the properties of BHs and compact objects belonging to four sub-populations of interest: isolated BHs from single-star origin, disrupted BHs of binary-star origin, main-sequence stars with a compact object companion, and double compact object mergers. 
We present the mass and center of mass (CoM) speed of BHs and compact binaries following the predictions of rapid population synthesis methods (Section \ref{sec:method}).
We present these populations generally and place them in the context of isolated BHs (Section \ref{sec:disc}), particularly \event.

\section{Binary Population Synthesis} 
\label{sec:method}
In order to simulate the demographics of BHs in the field we make use of the rapid binary population synthesis element of the COMPAS suite v02.27.05
\citep{2022ApJS..258...34R}.
The physical assumptions from this population have been chosen to match predictions of Galactic double neutron stars \citep{2018MNRAS.481.4009V}, double compact object mergers \citep[e.g.][]{2019MNRAS.490.3740N}, and Be X-ray binaries in the Small Magellanic Cloud \citep{2020MNRAS.498.4705V}.

We simulate $5\times 10^6$ binaries at representative metallicities.
These metallicities are the lowest metallicity models from our stellar evolution tracks \citep[$Z=0.0001$,][]{2000MNRAS.315..543H,2022ApJS..258...34R}, a representative metallicity for IZw18 \citep[$Z=0.0002$,][]{2015A&A...581A..15S}, $Z=0.001$, a representative metallicity for the Small Magellanic Cloud \citep[$Z=0.0021$,][]{2011A&A...530A.115B}, a representative metallicity for the Large Magellanic Cloud \citep[$Z=0.0047$,][]{2011A&A...530A.115B}, $Z=0.01$, a metallicity representative to the Sun \citep[$Z=0.0142$,][]{2009ARA&A..47..481A}, and a super-solar metallicity $Z=0.02$.
The details of the initial conditions and setup are presented in Table \ref{tab:COMPAS} in Appendix \ref{app:compas_setup}.
This data is available via \href{https://doi.org/10.5281/zenodo.6346443}{zenodo} \citep{team_compas_2022_6346444}.

COMPAS keeps track of the CoM and component speed of the  individual binary members, and resolves how are they modified when a supernova occurs. 
To do this, COMPAS follows Appendix B of \cite{2002ApJ...573..283P}.
If a supernova leads to a disruption of the binary, each element speed of the components of the binary is computed.
In the case that disruption occurs during the first supernova, the secondary star is followed throughout its remaining lifetime in order to define the stellar remnant and natal kick associated to it.
Currently, COMPAS does not follow the evolution of stellar mergers, and therefore they are ignored in this analysis (more information about the implication of stellar mergers in Section \ref{sec:disc}).

Here, we focus on the mass and CoM speed of systems that are comprised of at least one compact object.
Particularly, we center our attention on systems that will lead to BH formation.
We distinguish between isolated BHs and compact-object binaries.
Isolated BHs can be of single stellar origin ($\rm BH_{sin}$), or from a binary that becomes disrupted after a supernova ($\rm BH_{1,2}$ where the subscripts 1 and 2 to indicate that the progenitors of the now disrupted BH were initially the more and less massive star of the binary, respectively).
Compact-object binaries are those that form when the first supernova leads to NS or BH formation in a bound orbit.
For compact-object binaries, we only consider systems with a main sequence (MS) companion, either NS-MS or BH-MS binaries.
These systems are potential progenitors of X-ray binary sources, such as Be X-ray binaries \citep{2011Ap&SS.332....1R} or high-mass X-ray binaries \citep{2006csxs.book..623T}.
Finally, we consider double compact object mergers, either BH-NS, BH-BH, or NS-NS binaries.
For BH-NS systems, we do not distinguish which compact object formed first.
For double compact objects, we only consider those that will merge within a Hubble time via gravitational-wave emission \citep{1964PhRv..136.1224P}. 
We focus on double compact object mergers for two reasons: i) because they are the systems that will eventually become a single lens, and ii) because they are the systems that will be closer together and more likely to be confused with a single-lens source.
For isolated BHs, the mass and speed are those of the remnant after the supernova.
For binaries, the mass and speed correspond to the total mass and the CoM speed of the binary.

\begin{figure}
	\includegraphics[trim=24 7.5cm 2cm 7.5cm,clip,width=\columnwidth]{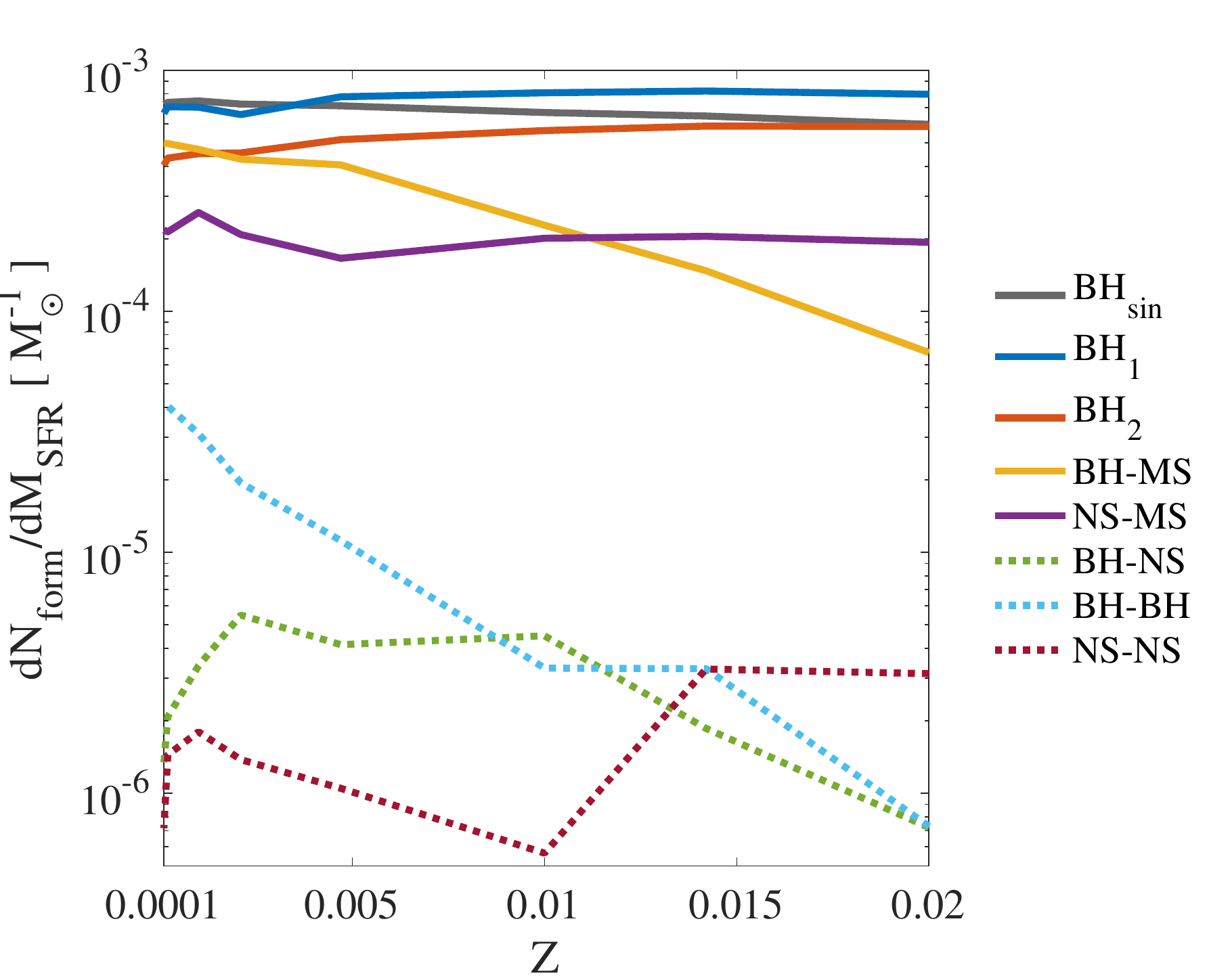}
	\caption{
	Number of systems of interest per unit star forming mass. The various yields are calculated  based on a population of $5 \times 10^6$ massive binaries per metallicity (Section \ref{sec:method}).
	BH$_{\rm{sin}}$ (solid grey) represents isolated black holes from single-star origin.
	BH$_1$ (solid blue) and BH$_2$ (solid orange) represent disrupted BHs from binary origin; the subscript 1 and 2 indicate that the progenitor is the initially more and less massive star, respectively.
	BH-MS (solid yellow) and NS-MS (solid purple) represent bound binaries, comprised of a compact object with a main sequence companion, just after the first supernova.
	Double compact objects that will merge within the Hubble time are shown as dotted lines: BH-NSs (green), BH-BH (cyan), and NS-NS (burgundy).
	The color scheme is the same in all figures.
	}
    \label{fig:yields}
\end{figure}

We classify and count every system from each of the aforementioned sub-populations and estimate their respective  yield, which is the number of systems of interest per unit star forming mass. 
To estimate the relative contribution of each channel, we need to correct for two things.
The first one is to include single stars in our population, which initially assumed 100\% binarity.
We do so by reusing the primary stars of the widest binaries ($a>80$ AU\footnote{This value results in $f_{\rm{bin}}\approx 0.8$ as specified later.}, where $a$ is the semi-major axis), and using their imparted natal kick, while ignoring the secondary and post-binary evolution (if any).
The initial separation distribution is assumed to be a log-uniform distribution, i.e., $p(a)\propto a^{-1}$, and therefore primaries from binaries between $100 \leq a/\rm{AU} \leq 1000$ will follow the initial mass function (IMF) and are likely to be non-interacting: i.e., component stars from wide binaries are likely to evolve as two effectively single stars.
We also consider a binary fraction, $f_{\rm{bin}}$, so that $N_{\rm{tot}} = N_{\rm{bin}} + N_{\rm{sin}} = (f_{\rm{bin}}+f_{\rm{sin}})\times N_{\rm{tot}}$, where $N_{\rm{tot}}$ is the total number of systems, $N_{\rm{bin}}$ is the total number of binaries, $N_{\rm{sin}}$ is the total number of single stars, and $f_{\rm{sin}}$ is the fraction of single-star systems\footnote{We assume that higher-multiplicity systems are comprised of an interacting inner binary, such as the ones from our synthetic population.}.
We calculate the number of single stars we need to account for following a binary fraction as $N_{\rm{sin}}= N_{\rm{bin}}(1-f_{\rm{bin}})/f_{\rm{bin}}$. We use $f_{\rm{bin}}=0.8$, which is adequate for stars with initial mass $M>10\ \rm{M_{\odot}}$ \citep{2017ApJS..230...15M}, which are potential BH progenitors.
The second thing we  correct for is the total mass of the population. 
Our simulation  includes only stars with mass $M \ge 5\ \rm{M_{\odot}}$, which follow the \cite{1955ApJ...121..161S} IMF.
We use the full \cite{2001MNRAS.322..231K} IMF to account for lower masses, assuming that all low-mass stars are single.
With these two corrections, we can quantify the yield per sub-population of interest as a function of metallicity (Figure \ref{fig:yields}).

In the simulated population (Figure \ref{fig:yields}), most isolated BHs ($\approx 8\times10^{-4}\ \rm{M_{\odot}^{-1}}$) come from the disruption of the primary BH after the first supernova, with similar yields at all metallicities.
The yield of single BHs and disrupted secondaries ($\approx 6\times10^{-4}\ \rm{M_{\odot}^{-1}}$) is similar at high metallicities ($Z\approx 0.02$).
However, the yield of single BHs increases at lower metallicities; at $Z\lesssim 0.0002$, they become similar to those of BHs from disrupted primaries.
The yield of $\rm BH_2$, which is comprised of the isolated BHs that originate from the disruption of the secondary, decreases as a function of metallicity ($\approx 4\times10^{-4}\ \rm{M_{\odot}^{-1}}$ at $Z=0.0001$).
The yield of NS-MS systems ($\approx 2\times10^{-4}\ \rm{M_{\odot}^{-1}}$) is roughly constant at all metallicities.
The yield of BH-MS systems is non-negligible at low metallicities ($\approx 5\times10^{-4}\ \rm{M_{\odot}^{-1}}$ at $Z=0.0001$) and decreases monotonically at high metallicities ($\approx 7\times10^{-5}\ \rm{M_{\odot}^{-1}}$ at $Z=0.02$).
Merging double compact objects are rare ($< 10^{-4}\ \rm{M_{\odot}^{-1}}$ at all simulated metallicities).
The decreasing yields as a function of metallicity of BH-MS and BH-BH binaries are correlated with the corresponding increased mass-loss rates via stellar winds at high-metallicity environments \citep[e.g.,][]{2019MNRAS.490.3740N}.

\begin{figure}
	\includegraphics[trim=1cm 7.5cm 1cm 7.5cm,clip,width=\columnwidth]{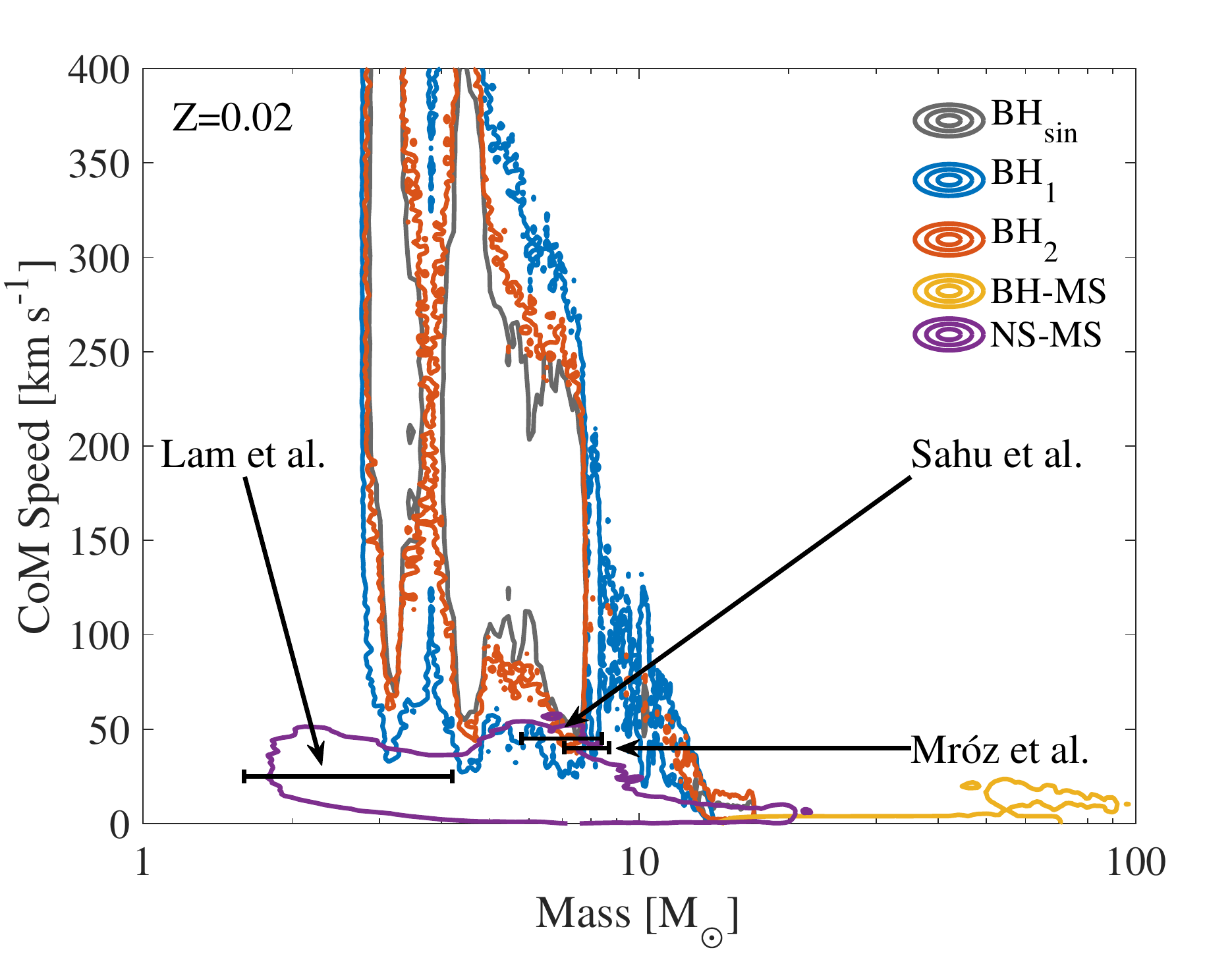}
    \includegraphics[trim=1cm 7.5cm 1cm 7.5cm,clip,width=\columnwidth]{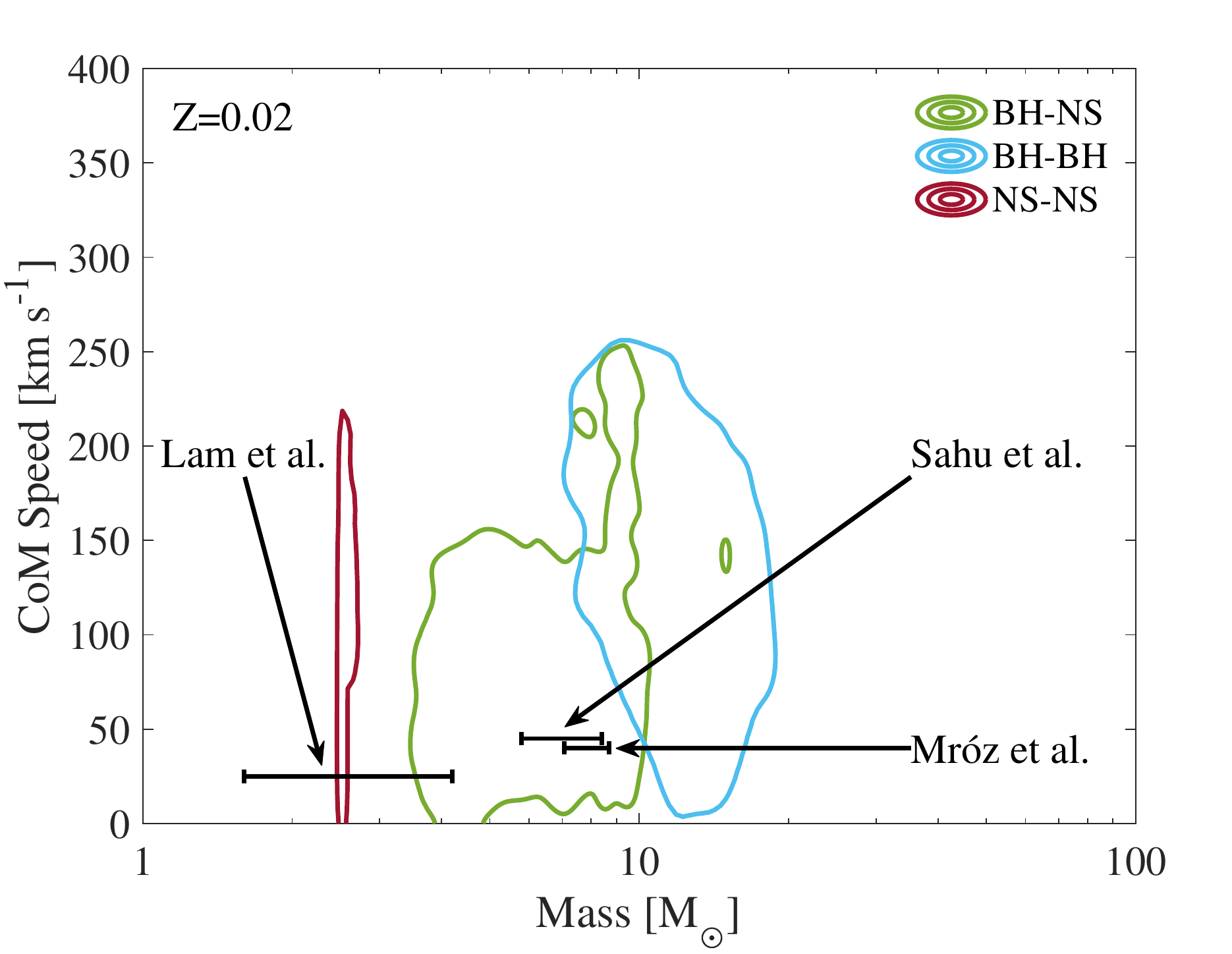}
    \caption{
    Kernel density estimation showing the \confidence confidence interval regions for the sub-population of interest (Figure \ref{fig:yields} and Section \ref{sec:method}) at $Z=0.02$.
    The speed corresponds to the CoM velocity magnitude.
    The mass corresponds to the BH or total binary mass of the system, depending whether or not the system is isolated or bound.
    We show the different reported lens mass of \event\ as error bars and the transverse space velocity, according to \cite{2022ApJ...933L..23L}, \cite{2022ApJ...933...83S} and the more recent analysis from \cite{2022ApJ...937L..24M}.
    The transverse space velocity value from \cite{2022ApJ...933L..23L} is an upper limit.    
    Top: sub-populations of isolated BHs (BH$_{\rm{sin}}$, BH$_1$, BH$_2$) and compact-object binaries (BH-MS, NS-MS).
    Bottom: sub-populations of double compact object mergers (BH-NS, BH-BH, NS-NS).
    }
    \label{fig:mass_and_speed}
\end{figure}

After we have identified each sub-population of interest, we estimate their respective mass and CoM speed probability density function using kernel density estimation (KDE).
We calculate an adaptive KDE based on linear diffusion processes as presented in \cite{10.1214/10-AOS799}. 
The kernel for such KDE is assumed to be Gaussian and the bandwidth parameters are chosen optimally without the need of arbitrary choices or ``rules-of-thumb" \citep{10.1214/10-AOS799}.
We use the KDE to define \confidence confidence-interval \textit{contours} for each sub-population (Figure \ref{fig:mass_and_speed}).

\begin{figure*}
	\includegraphics[trim=1cm 7.5cm 1cm 7.5cm,clip,width=0.5\linewidth]{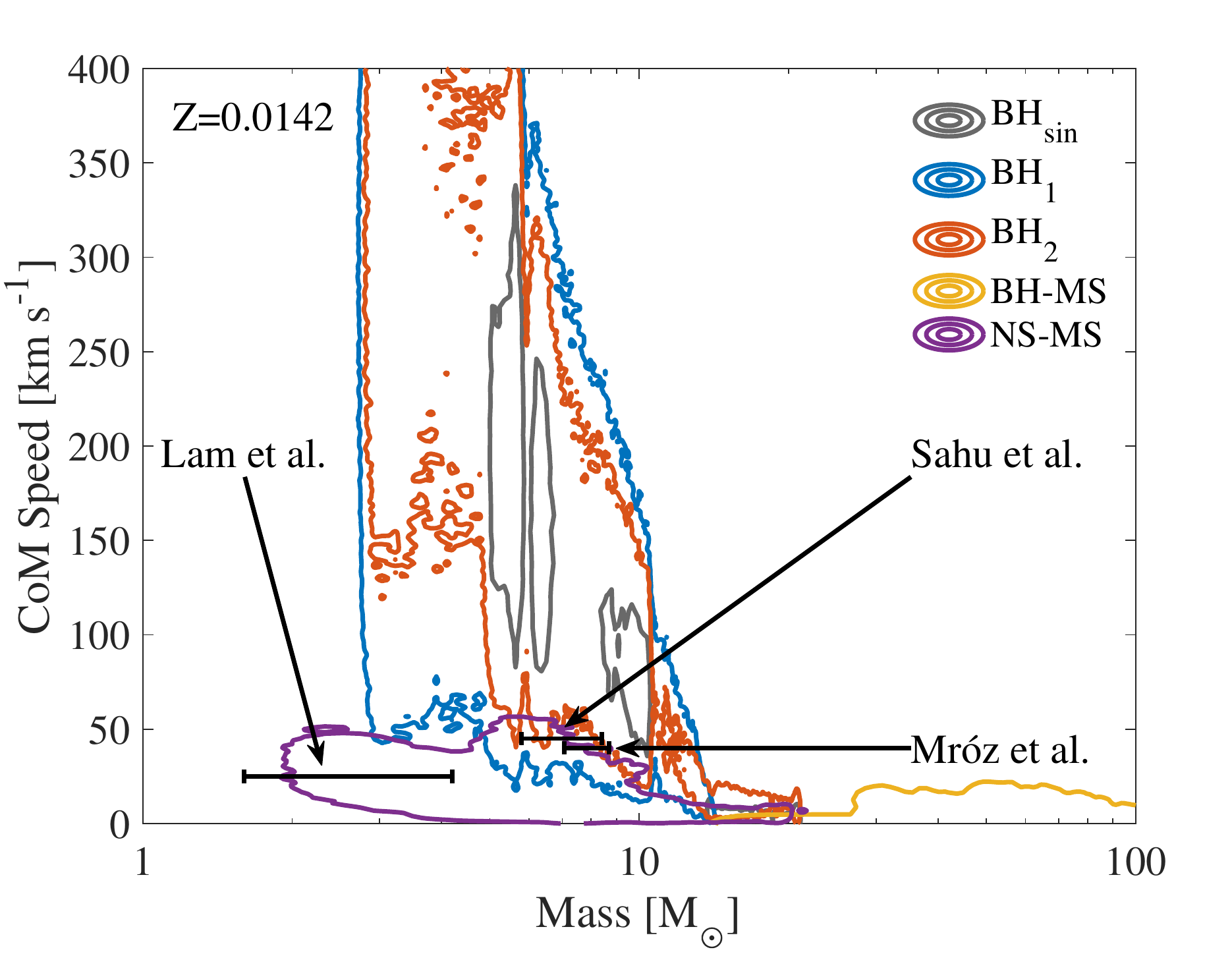}
    \includegraphics[trim=1cm 7.5cm 1cm 7.5cm,clip,width=0.5\linewidth]{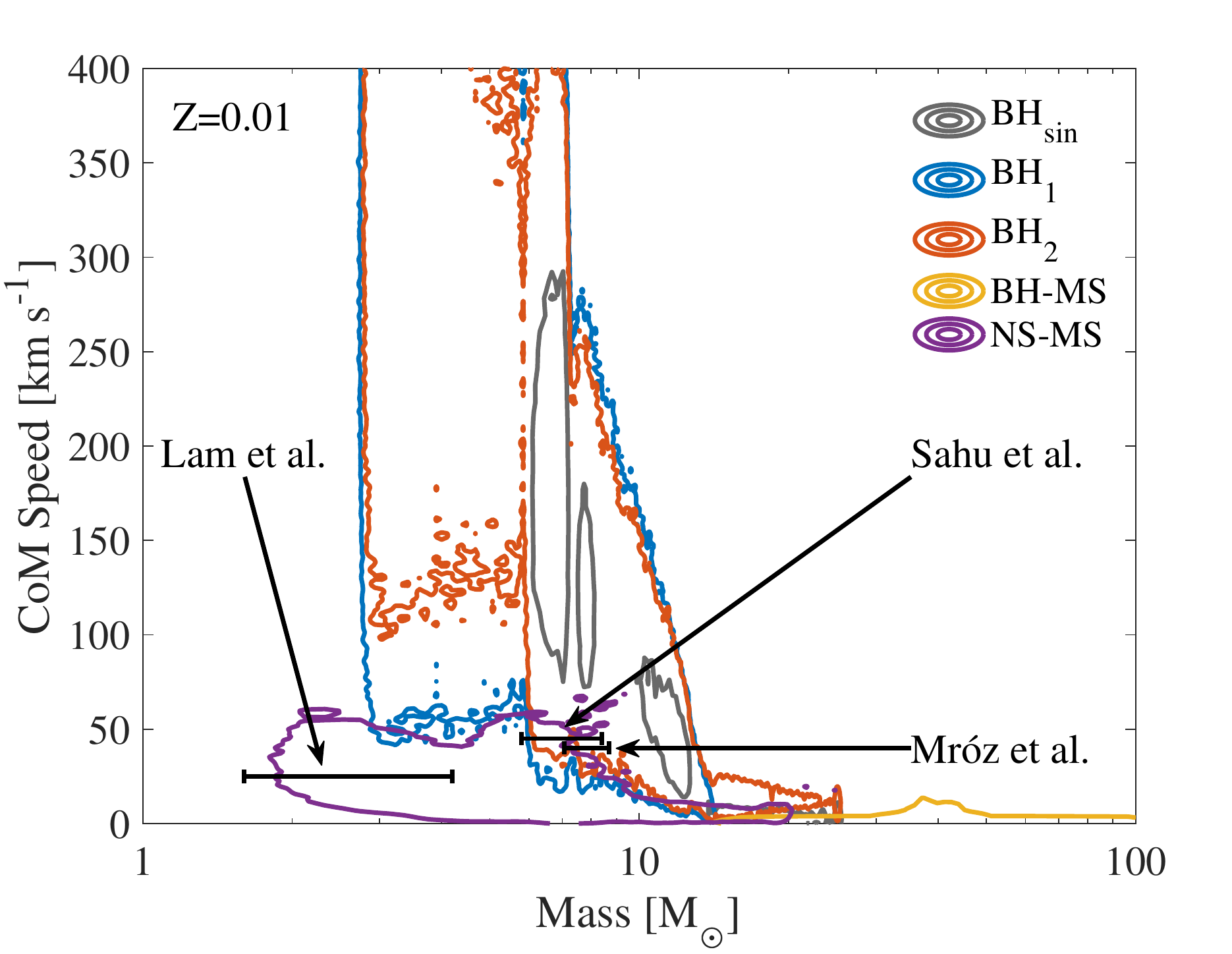}
    \includegraphics[trim=1cm 7.5cm 1cm 7.5cm,clip,width=0.5\linewidth]{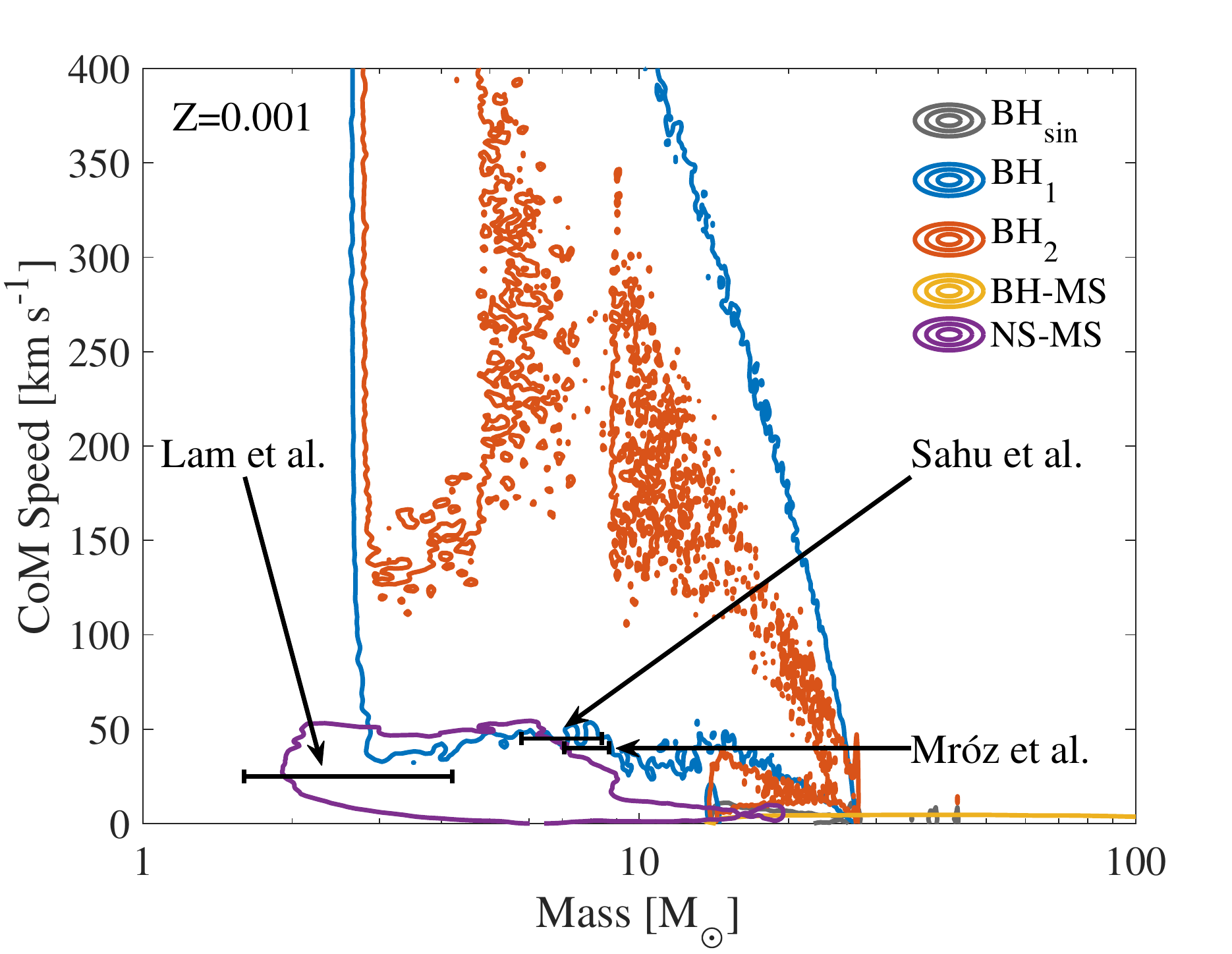}
    \includegraphics[trim=1cm 7.5cm 1cm 7.5cm,clip,width=0.5\linewidth]{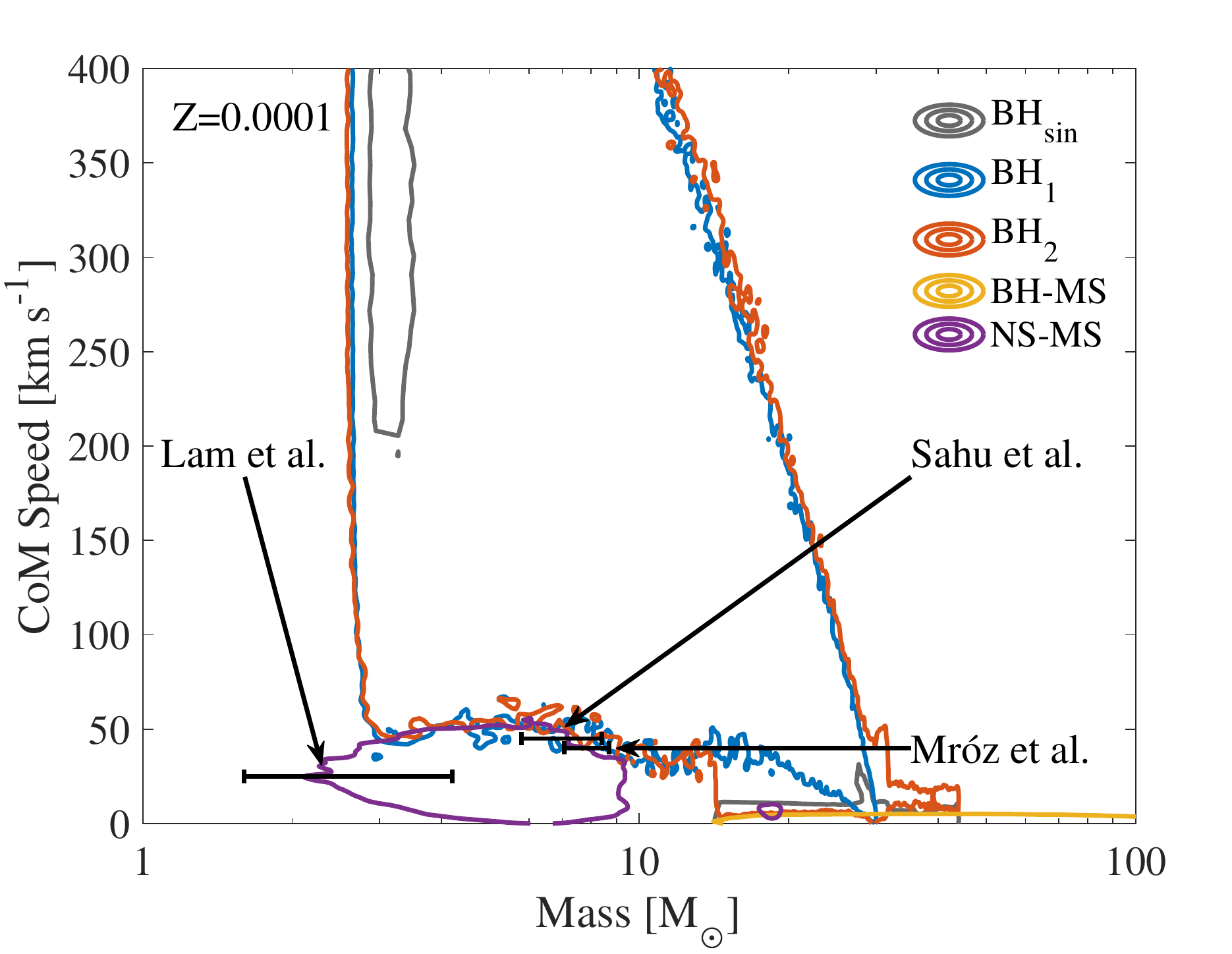}
	\caption{
	KDE contours showing the \confidence confidence interval regions for the isolated BH and compact-binary sub-populations at $Z=\{0.0142,0.01,0.001,0.0001\}$.
	The bulk of the single BH population at $Z=0.001$ (bottom-left panel) are low-speed ($\approx 0\  \rm{km\ s^{-1}}$) and with masses $> 10\ \rm{M_{\odot}}$, and therefore barely visible in this KDE.
	For more details, see Section \ref{sec:method} and Figure \ref{fig:mass_and_speed}.
	}
    \label{fig:mass_and_speed_all_metallicities_top}
\end{figure*}

\begin{figure*}
	\includegraphics[trim=1cm 7.5cm 1cm 7.5cm,clip,width=0.5\linewidth]{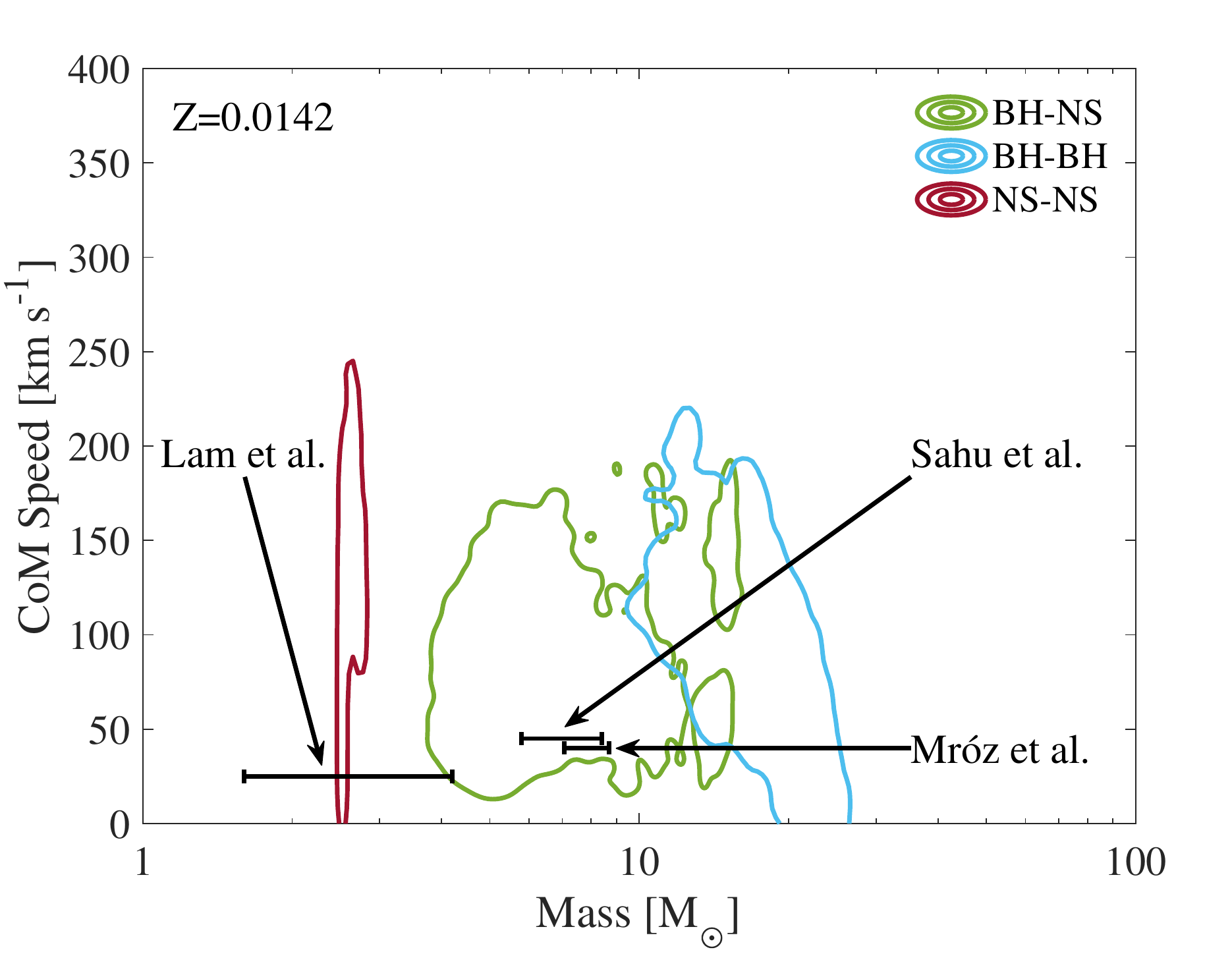}
    \includegraphics[trim=1cm 7.5cm 1cm 7.5cm,clip,width=0.5\linewidth]{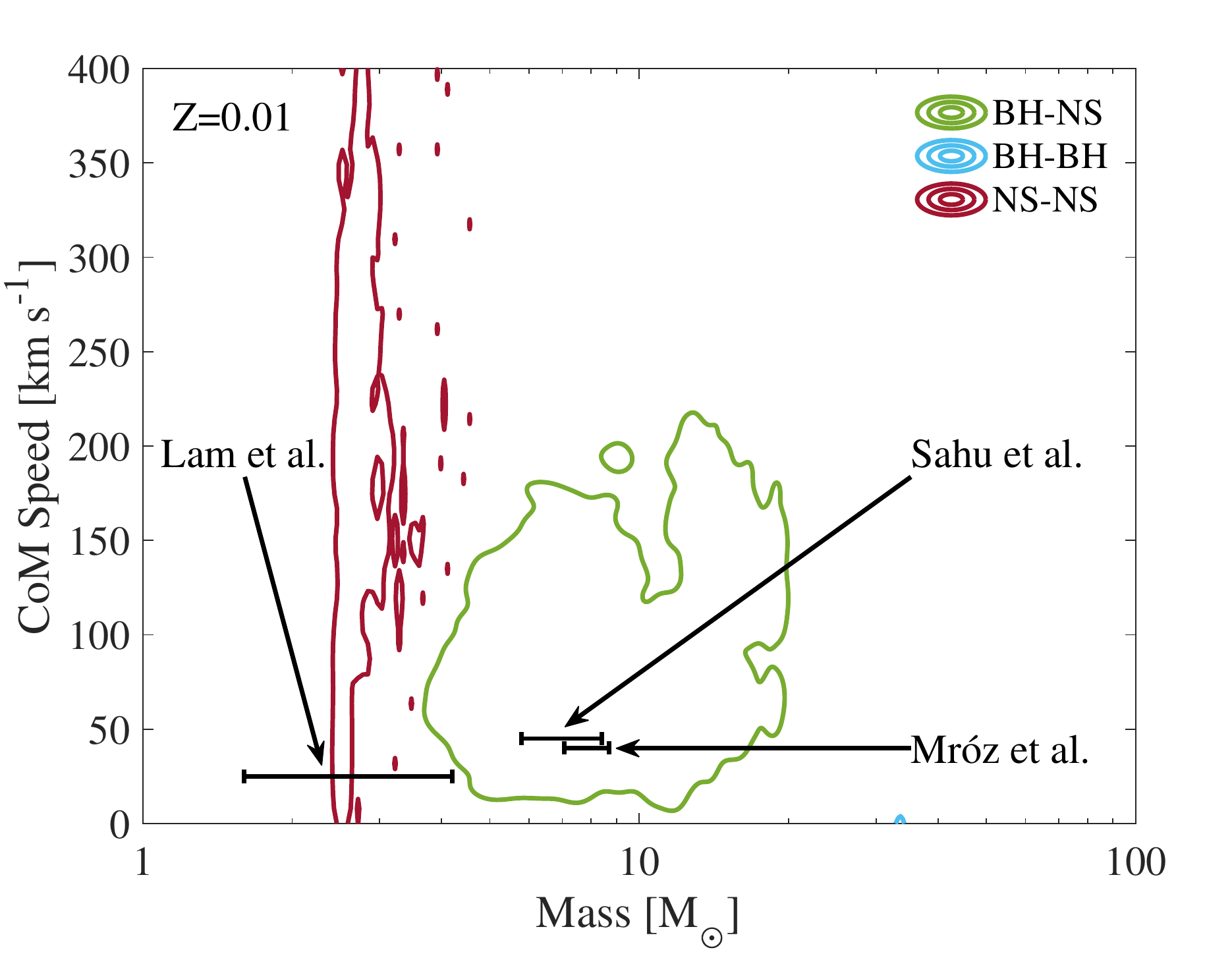}
    \includegraphics[trim=1cm 7.5cm 1cm 7.5cm,clip,width=0.5\linewidth]{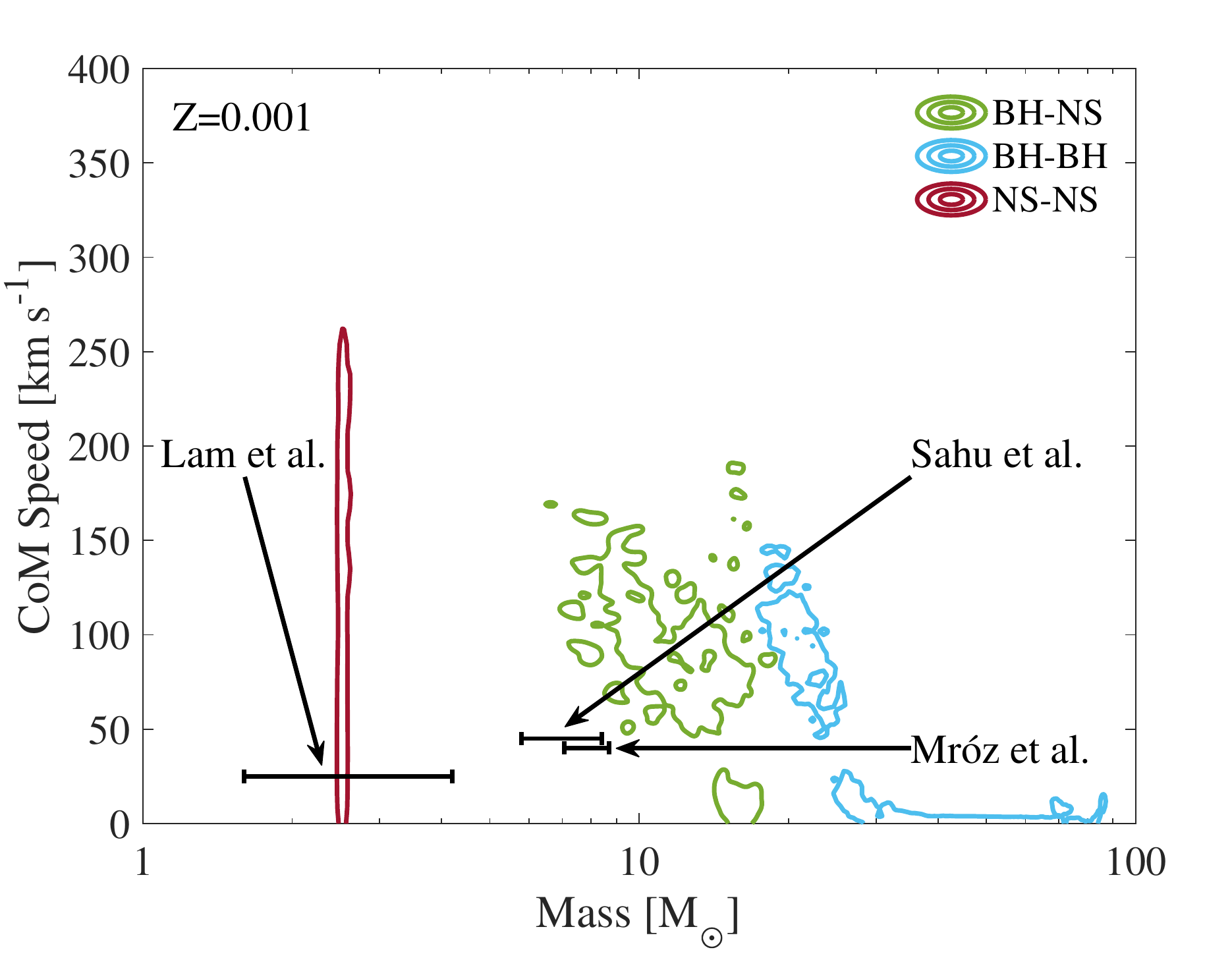}
    \includegraphics[trim=1cm 7.5cm 1cm 7.5cm,clip,width=0.5\linewidth]{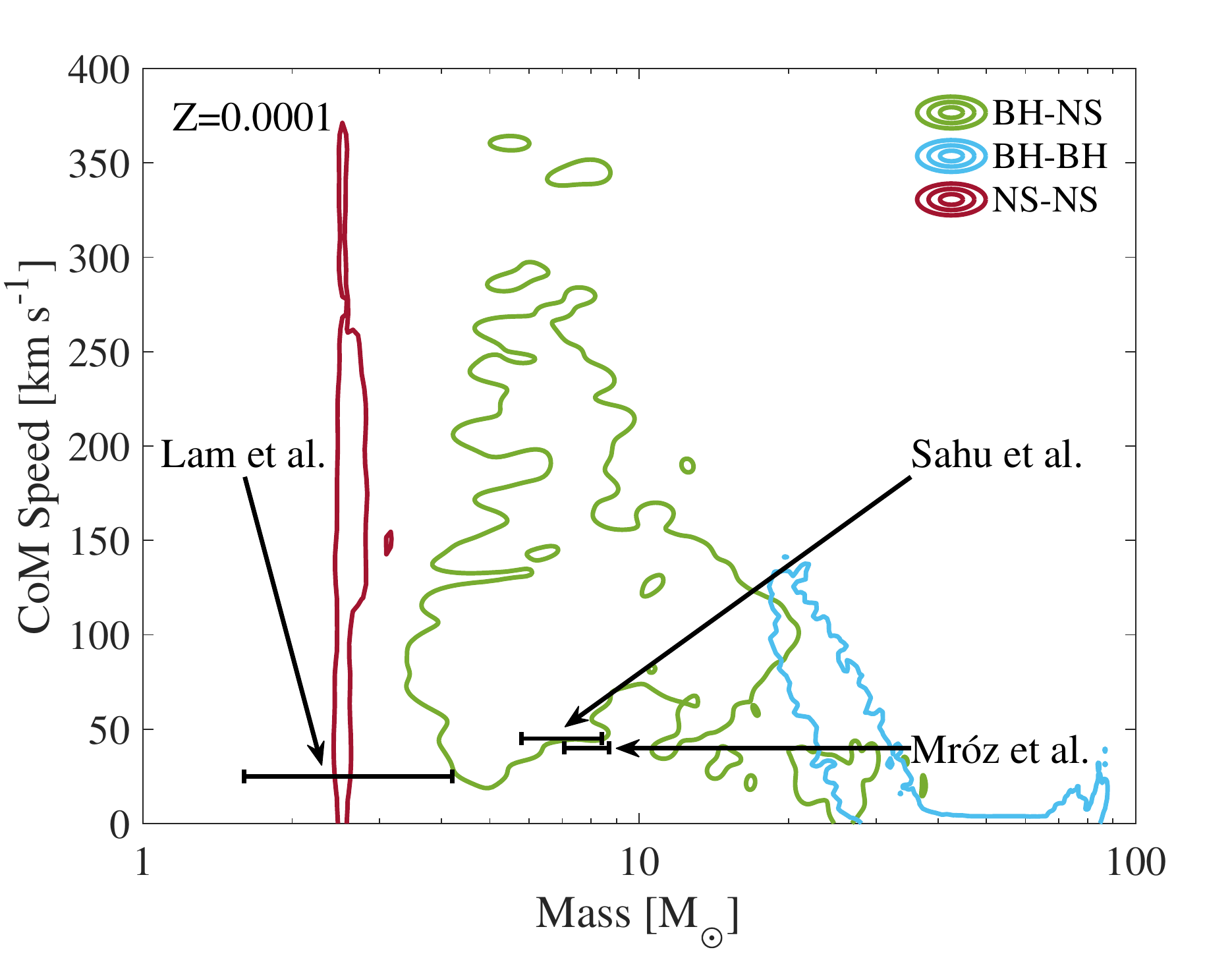}
	\caption{
	KDE contours showing the \confidence confidence interval regions for double compact object merger sub-populations at $Z=\{0.0142,0.01,0.001,0.0001\}$.
	The bulk of the BH-BH population at $Z=0.01$ (top-right panel) are low-speed ($\approx 0\ \rm{km\ s^{-1}}$) and with masses around $\gtrsim 30\ \rm{M_{\odot}}$, and therefore barely visible in this KDE.
	For more details, see Section \ref{sec:method} and Figure \ref{fig:mass_and_speed}.
	}
    \label{fig:mass_and_speed_all_metallicities_bot}
\end{figure*}

In Figure \ref{fig:mass_and_speed}, we present these contours for the highest metallicity population ($Z=0.02$). The contours of the populations at other lower metallicities are presented in Figures \ref{fig:mass_and_speed_all_metallicities_top} and \ref{fig:mass_and_speed_all_metallicities_bot}.
The different inferred values from \event, as reported by \cite{2022ApJ...933...83S} and \cite{2022ApJ...933L..23L}, lead to different interpretations of the results.
The more massive lens mass and higher transverse space velocity values, as reported by \cite{2022ApJ...933...83S}, lie within the contours of the BH$_1$, BH$_2$, and NS-MS, and BH-NS sub-populations.
However, the less massive and lower transverse space velocity values, as reported by \cite{2022ApJ...933L..23L}, lie only within the contour of the NS-MS sub-population, and marginally within the contour of the binary-origin BH$_1$, NS-NS, and BH-NS sub-populations.
None of the reported values for \event\ lie within the contour of neither the BH$_{\rm{sin}}$, the BH-MS, nor the BH-BH sub-populations.

\begin{figure}
	\includegraphics[trim=0.5cm 7.5cm 2.0cm 7.5cm,clip,width=\columnwidth]{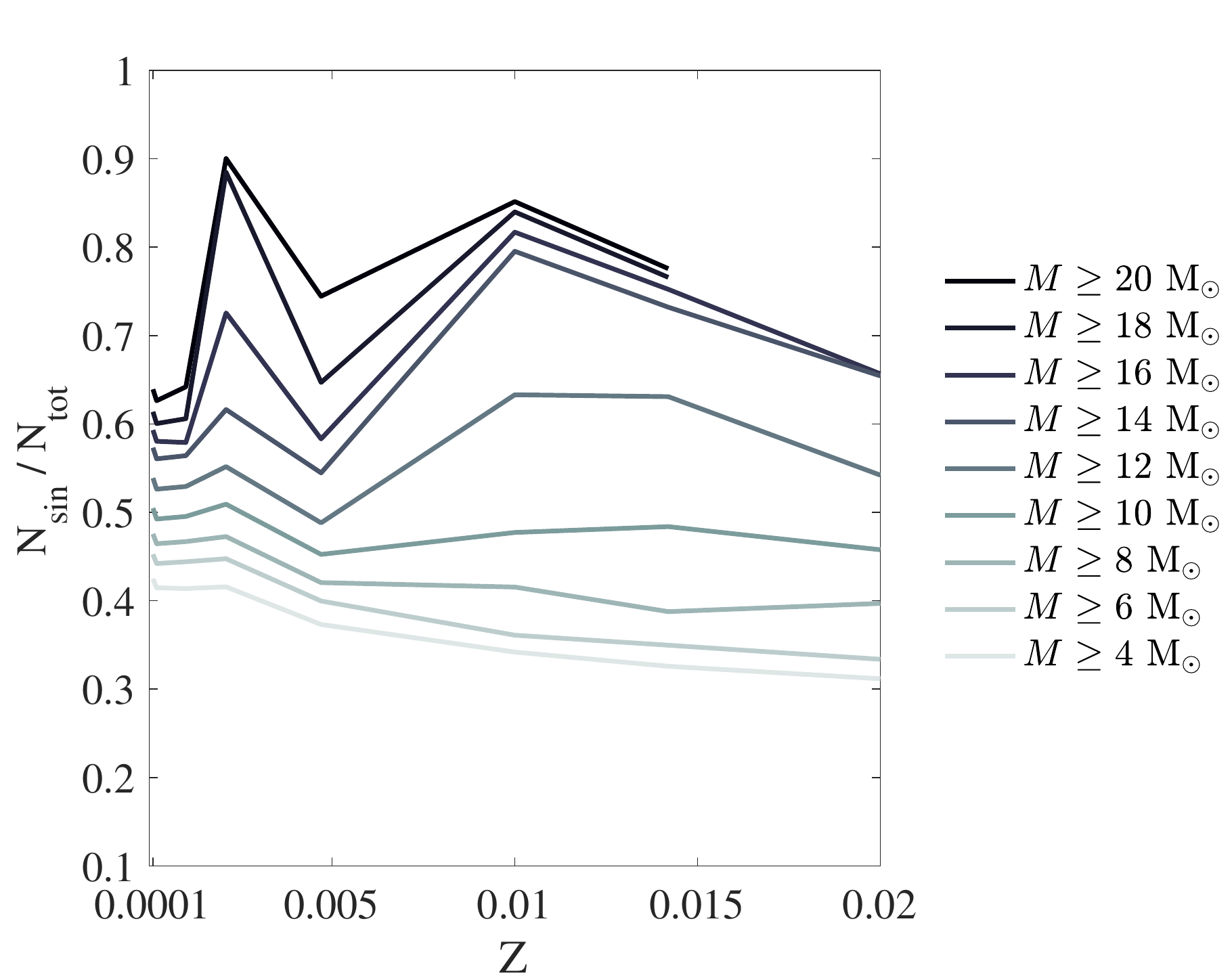}
    \caption{
    Fraction of isolated BHs of single origin ($N_{\rm{sin}}/N_{\rm{tot}}$), for masses above the threshold mass $M$, as a function of the simulated metallicities.
    Isolated high-mass ($\gtrsim\mass\ \rm{M_{\odot}}$) BHs are predominantly of single-star origin.
    The darker lines are discontinuous at high metallicities because stellar winds remove enough mass that there are not BHs with mass $\gtrsim20\ \rm{M_{\odot}}$ in our model (Appendix \ref{app:compas_setup}).
    }
    \label{fig:fractions}
\end{figure}

Finally, we quantify the fraction of BHs from single origin (Figure \ref{fig:fractions}).
Approximately 70\% of low-mass ($\leq \mass\ \rm{M_{\odot}}$) isolated BHs are from binary origin.
In contrast, most high-mass ($> 10\ \rm{M_{\odot}}$) isolated BHs are from single origin.
Moreover, the more massive the BH, the most likely it is to be from single origin (up to $\approx 80-90$\% at some metallicities).
If we assume that \event\ is a BH, or a massive ($>1.6\ \rm{M_{\odot}}$) NS, it is more likely than it was originated from a binary-star system than from a single star.

\begin{figure*}
	\includegraphics[trim=1cm 7.5cm 1cm 7.5cm,clip,width=0.5\linewidth]{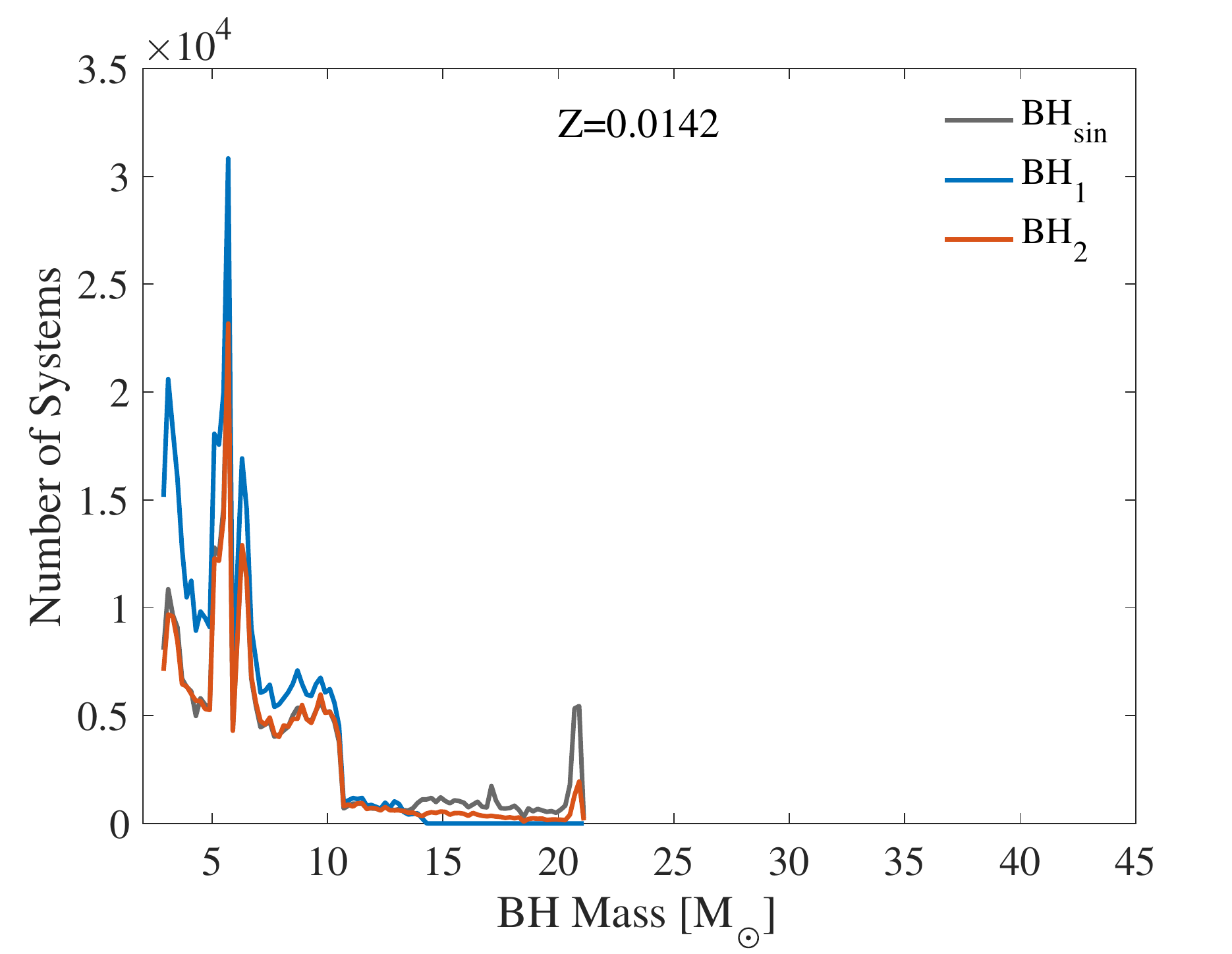}
    \includegraphics[trim=1cm 7.5cm 1cm 7.5cm,clip,width=0.5\linewidth]{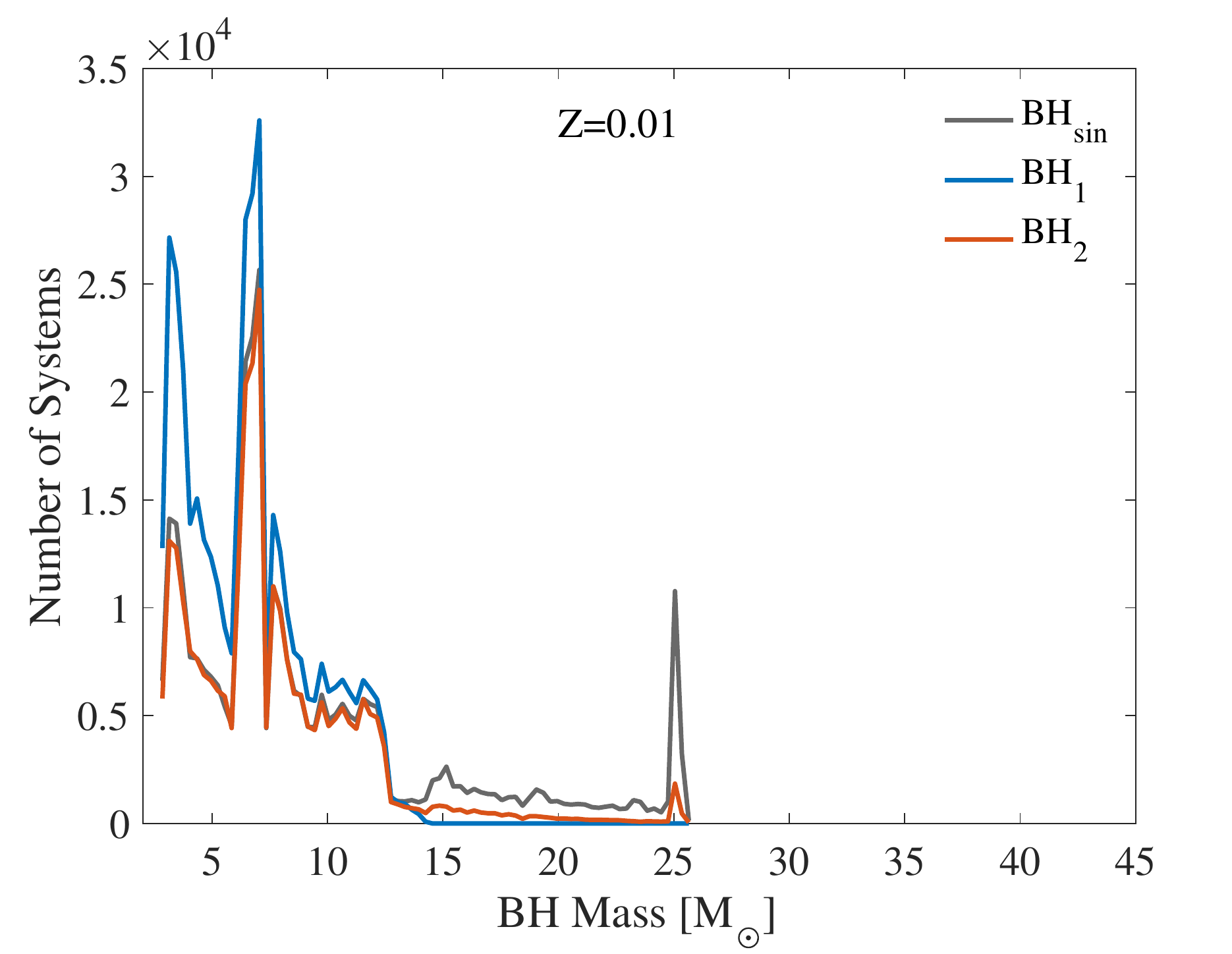}
    \includegraphics[trim=1cm 7.5cm 1cm 7.5cm,clip,width=0.5\linewidth]{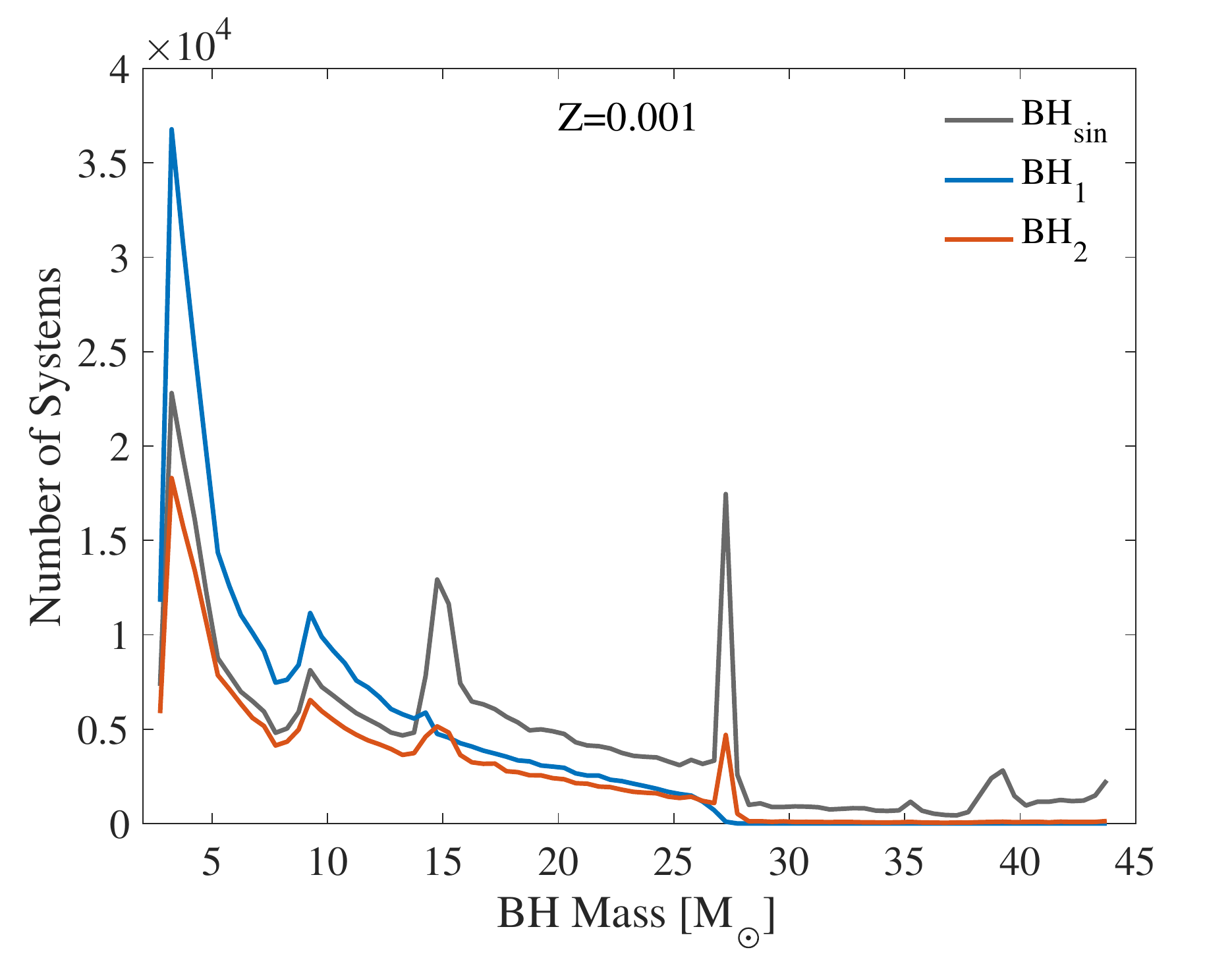}
    \includegraphics[trim=1cm 7.5cm 1cm 7.5cm,clip,width=0.5\linewidth]{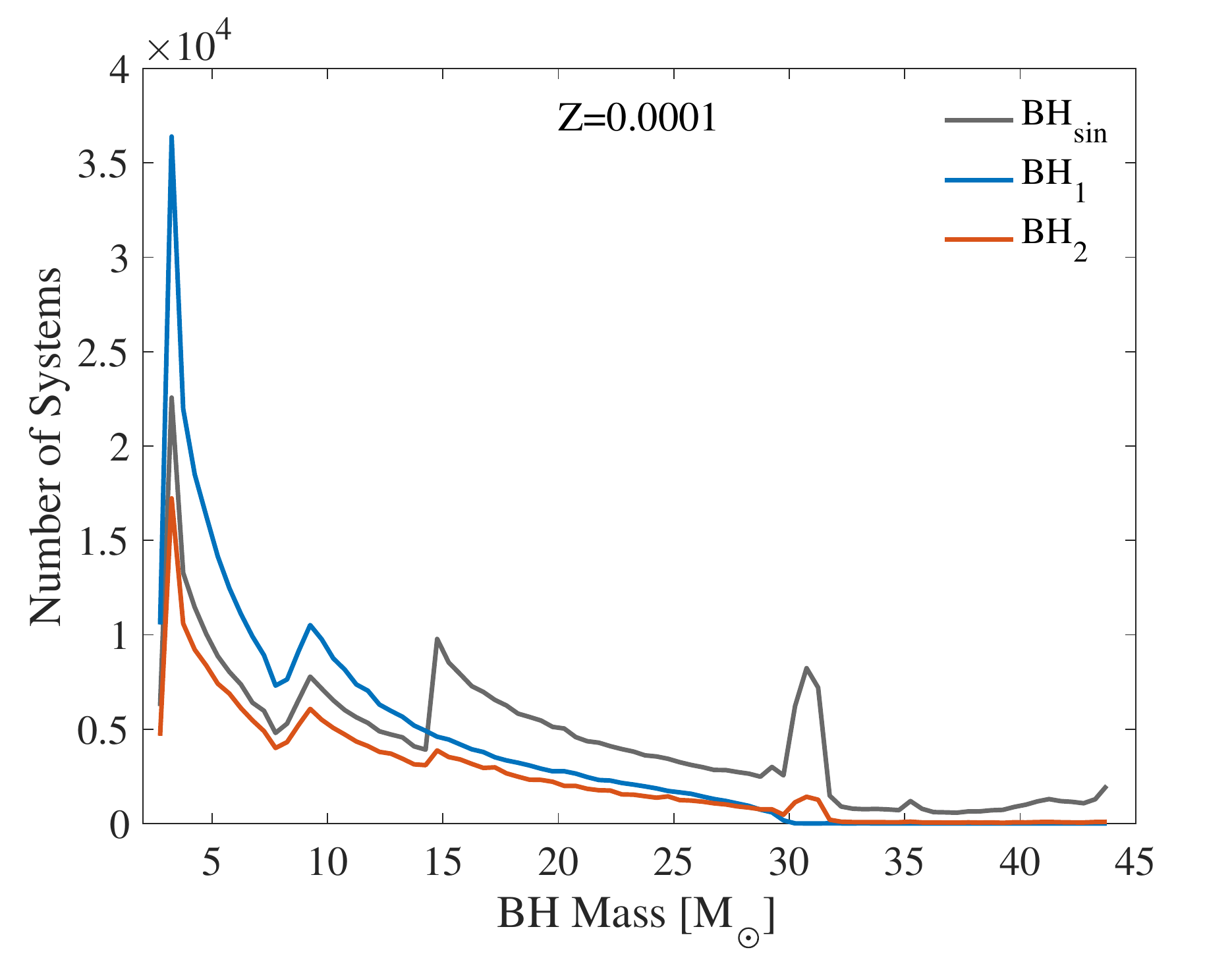}    
    \caption{
    BH mass distribution at at $Z=\{0.0142,0.01,0.001,0.0001\}$.
    The number of isolated BHs from single origin increase as a function of mass for all metallicities.
    This effect is enhanced at lower metallicities.
    Low-mass ($\lesssim 10\ \rm{M_{\odot}}$) BH lenses can be used to explore binary evolution, while high-mass ($\gtrsim 10\ \rm{M_{\odot}}$) BH lenses can be used to understand single stellar evolution (Section \ref{subsec:learn}).
    }
    \label{fig:mass_distribution_all_metallicities}
\end{figure*}

\begin{figure}
	\includegraphics[trim=1cm 7.5cm 1cm 7.5cm,clip,width=\columnwidth]{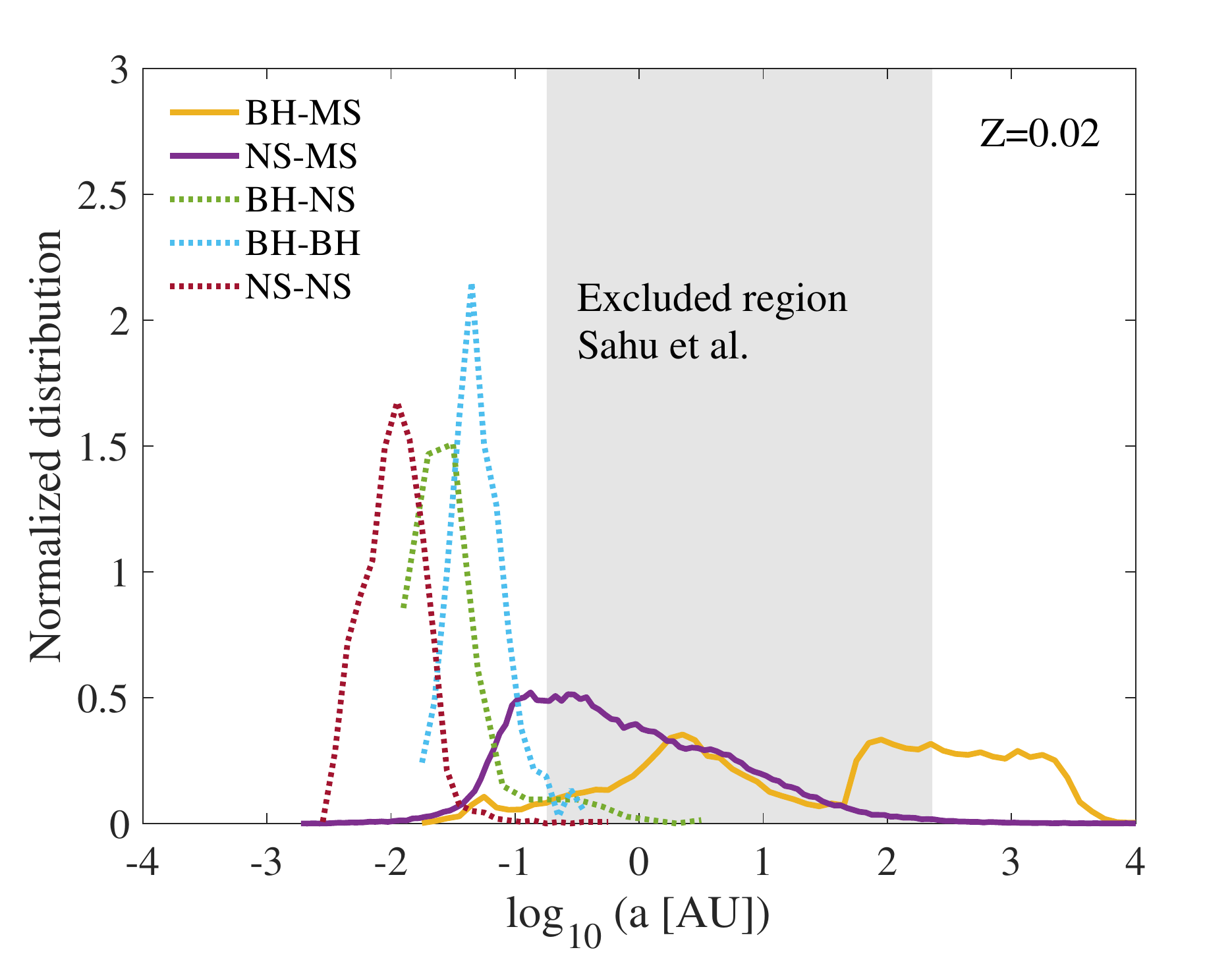}
    \caption{
    Normalized distributions of the semi-major axis of all binary sub-populations.
    Compact objects with a main sequence companion (NS-MS, BH-MS) are shown in solid lines.
    NS-MS and BH-MS are X-ray binary candidates, particularly if they are in close orbits or eccentric enough than their periastron passage triggers X-ray emmitting processes.
    Double compact object mergers (BH-NS, BH-BH, and NS-NS) are shown in dotted lines.
    They are in very short periods as we are only considering those that will merge within the Hubble time.
    The grey shaded region denotes the excluded region of binaries, or a companion to the lens mass of \event, as derived by \cite{2022ApJ...933...83S}. 
    }
    \label{fig:separation}
\end{figure}

\section{Discussion} 
\label{sec:disc}

\subsection{Summary} 
We present the landscape of isolated BHs from massive stellar populations in the context of microlensing event \event.
We did this using the rapid population synthesis element of the COMPAS suite.
Our model suggests that \event\ is an isolated BH of binary origin.
We find that, in addition to single and disrupted BHs, bound binaries can be a non-neglibile source of lenses in astronomical populations, both Galactic and extragalactic.
We find that most low-mass BHs are from binary origin and most high-mass BHs are from single origin, and that there is a distinctive population of such systems that can shed light on the origin of massive binaries and single massive stars in the Milky Way.

\subsection{Microlensing neutron stars} \label{subsec:neutron_stars}
The fate of stellar remnants depend strongly on their properties at birth.
NS remnants are believed to experience a natal kick at formation \citep[e.g.][]{2004ApJ...612.1044P}.
In our COMPAS model, the natal kicks after core-collapse supernovae follow the isolated pulsar velocity distribution \citep{2005MNRAS.360..974H}. 
Alternatively, electron-capture supernovae have been predicted to receive low natal kicks \citep{2004ApJ...612.1044P} and leave behind a NS with mass $\lesssim 1.3\ \rm{M_{\odot}}$ \citep{1984ApJ...277..791N}.
Therefore, our COMPAS model assumes a bimodal, high-mode ($\sigma=265\ \rm{km\ s^{-1}}$) and low-mode ($\sigma=30\ \rm{km\ s^{-1}}$), kick distribution \citep{2022ApJS..258...34R}.
The low-mode kick is only pertinent for the NS-MS, BH-NS, and NS-NS sub-populations, and can only propagate into the isolated BH distribution via mergers of these sub-populations (Section \ref{subsec:DCOs}).

While the low transverse velocity of \event\ could be associated to low-kick magnitude supernova, modulo the complexity of kinematics, the reported mass ($>1.6\ \rm{M_{\odot}}$) is too heavy \citep{2022ApJ...933L..23L} to be associated with a canonical electron-capture supernova.
\cite{2022ApJ...933L..23L} reported 4 additional events to \event, including MB09260. 
MB09260 is reported with a 0.38:0.44:0.14 probability of being a white dwarf, NS, and BH, respectively.
Based solely on the reported (global median) lens mass of $1.37\ \rm{M_{\odot}}$, similar to the baryonic mass of the ONeMg core leading to electron-capture supernovae \citep{1984ApJ...277..791N}, we consider MB09260 as a more appropriate candidate of an electron-capture supernova \citep[but see][for a discussion on the mass uncertainties]{2022ApJ...933L..23L}.
Alternatively, it could also be that the progenitor of MB09260 experienced stripping at some point in its evolution, a process which has been associated to reduced natal kicks \citep{2021ApJ...920L..37W}.

\subsection{Microlensing black holes}
Stars more massive than $\approx 20\ \rm{M_{\odot}}$ are thought to be the progenitors of BHs.
BH progenitors are more rare and less likely to be single ($f_{\rm{sin}}<0.1$ for stars with $M \gtrsim 20\ \rm{M_{\odot}}$ according to \citealt{2017ApJS..230...15M}).
Additionally, the natal kick associated with massive BH remnants is believed to be smaller in magnitude with respect to that of NS progenitors \citep[e.g.][]{2012ApJ...749...91F}.
In COMPAS, the natal kick BHs receive is drawn from the aforementioned high-mode distribution (Section \ref{subsec:neutron_stars}), but it is usually damped depending on the amount of fallback \citep{2012ApJ...749...91F}.
However, either the (reduced) natal kick or the amount of mass lost during BH formation might lead to a disruption of the binary.
The disruption of the binary after the first supernova can lead to a population of low-velocity ($\lesssim 30\ \rm{km\ s^{-1}}$) massive \textit{walkaway} stars \citep{2019A&A...624A..66R}, some of which are BH progenitors.
Overall, pre-supernova mass transfer onto a companion will lead to less massive stars at core-collapse \citep[e.g.][]{2021A&A...645A...6Z}, and likely reduced natal kicks.
This mass transfer event is also the reason why high-mass BHs are preferentially from single origin: when mass transfer occurs in massive binaries, it can lead to a mass redistribution where the component masses become less massive than the originally more massive star.
This mass redistribution depends on the mass of the donor and the accretor, and on how conservative the mass transfer episode is (Appendix \ref{app:compas_setup}).
Even for fully-conservative mass transfer, most configurations lead to a mass redistribution that results in post-mass-transfer stellar companions less massive than the originally more massive star.
For non-conservative mass transfer, this effect is only enhanced.
As a summary, for isolated BHs, the binary BH mass distribution is steeper than the single BH mass distribution (Figure \ref{fig:mass_distribution_all_metallicities}).

There is an estimated number of $\approx 10^8-10^9$ isolated BHs in the Galaxy \citep{1983bhwd.book.....S,1996ApJ...457..834T}.
\cite{2019ApJ...885....1W} show that the isolated BH population is dominated by isolated BHs of binary origin \citep[ also][]{2004ApJ...611.1068B,2020A&A...638A..94O}.
Moreover, \cite{2019ApJ...885....1W} accounts for stellar mergers, and conclude that the number of isolated BHs from mergers and disrupted binaries is comparable.
Our model does not include post-stellar-merger evolution, and therefore we are likely underestimating the amount of isolated BHs of binary origin; however, our model also overestimates the initial binary fraction ($f_{\rm bin} \approx  0.8$) with respect to observations ($f_{\rm bin} \approx  0.7$ according to \citealt{2012Sci...337..444S}).

\subsection{Surviving compact objects in binary systems}
\label{subsec:binaries}
X-ray binaries are a distinctive population of interacting MS stars with a NS or a BH companion \citep[e.g.,][]{2006csxs.book..623T,2011Ap&SS.332....1R}.
Our results focus on the moment of the formation of potential X-ray binary progenitors.
However, we do not follow the evolution of such systems, nor classify those that will emit X-rays and those that won't.
Particularly, our total mass estimates are upper limits, as we expect some systems to lose mass via stellar winds and mass transfer episodes.
Therefore, while there CoM velocity will not significantly change until the moment of the supernova, the total mass will likely decrease.
Some of these binaries will be wide, non-interacting ($>100$ AU, Figure \ref{fig:separation}), and therefore won't emit X-rays.
However, they might be observable with the Gaia mission \citep{2016A&A...595A...1G} that has been proposed to hunt for BHs \citep[e.g.,][]{2017MNRAS.470.2611M,2017ApJ...850L..13B,2018ApJ...861...21Y,2020ApJ...905..134W,2022A&A...658A.129J,2023ApJ...944..146A}.
Some MS-NS and MS-BH binaries might become giant star binaries \citep{2019Sci...366..637T} and could eventually experience a common-envelope episode \citep[e.g.][for a review]{2013A&ARv..21...59I}, which will further strip the stars, leading them to a close ($\lesssim 0.1$ AU) orbit configuration.
These close, stripped binaries might be difficult to detect \citep[e.g.][]{2018A&A...615A..78G}, as their luminosity is dependent on their mass and the amount of hydrogen left on their envelope after the stripping episode \citep[e.g.][]{2022MNRAS.511.2326V}.
In the case of \event, the presence of any companion more massive than 10\% of the identified
lens ($\approx 7.1\ \rm{M_{\odot}}$) has been excluded within 230 AU \citep{2022ApJ...933...83S} and there is no evidence for a massive, luminous main-sequence star companion at a larger separation.

\subsection{Double compact object binaries}
\label{subsec:DCOs}
Double compact objects detectable by ground-based gravitational-wave observatories need to be in close ($\lesssim 0.1$ AU) orbits in order to merge within the age of the Universe (Figure \ref{fig:separation}).
In such close orbits, the compact binary can be confused with a single lens (Section \ref{subsec:microlensing}). 
However, such configurations are extremely rare (Figure \ref{fig:yields}).
A double compact object merger leads to the formation of a single, rotating compact object, most likely a BH.
During inspiral, and mostly throughout the merger, mass-energy is radiated in the form of gravitational-waves, and therefore the post-merger mass is less than the sum of the component BH masses \citep[e.g.][]{2008PhRvL.100o1101B}.
Moreover, depending on the masses and spins of the component compact objects, the coalescence can lead to a gravitational-wave recoil kick for the post-merger remnant.
Non-spinning compact objects, likely representative of the field population \citep{2019ApJ...881L...1F} as well as those from the COMPAS simulation, will get a recoil kick in direction of the orbital plane with a magnitude as high as $\approx 175\ \rm{km\ s^{-1}}$ \citep{2007PhRvL..98i1101G}, and close to $\approx 0\ \rm{km\ s^{-1}}$ for either extreme mass ratios or mass ratios close to unity.
Therefore, we expect a double compact object merger to decrease the mass and slightly change the velocity with respect to the contours presented in the bottom panel of Figure \ref{fig:mass_and_speed}.

\subsection{What can we learn from black hole lenses}
\label{subsec:learn}
A population of BH lenses in the Milky Way can be used to explore the physics of massive stars.
\cite{2020ApJ...889...31L} performed a single-star single-metallicity population synthesis study, which accounts for selection effects, to suggest that the BH present-day mass function, BH multiplicity, and BH kick velocity distribution can be constrained using the future Nancy Grace Roman Space Telescope microlensing survey \citep{2015arXiv150303757S}.
That study didn't incorporate stellar binaries, and therefore is likely to be biased towards a shallower BH present-day mass function.

Shortly after we submitted this manuscript, \cite{2022ApJ...930..159A} explored how microlensing detections can be used to measure BH natal kicks, particularly in the context of \event.
Their method relies on backwards integration to trace the origin of compact objects by measuring their peculiar velocity and inferring the natal kick the object received.
In contrast, we have forward modeled a stellar population and calculated the velocities of isolated compact object lenses.
These methods are complementary and we reach a similar conclusion regarding mass-velocity configurations.

Here we have distinguished between a low- and a high- mass population of isolated BH lenses.
Low-mass ($\lesssim 10\ \rm{M_{\odot}}$) isolated BHs are more likely from binary origin, and therefore the physics of mass transfer, stripping, and orbital dynamics can be probed with these lenses.
Additionally, they can be used to explore supernovae, particularly the transition from neutron star to BH formation, and the \textit{so-called} first BH mass gap \citep[e.g.][and references therein]{2020A&A...636A..20W}.
High-mass ($\gtrsim 10\ \rm{M_{\odot}}$) isolated BHs are more likely from single origin, and therefore useful to explore the physics of supernova \citep[e.g.][]{2012ApJ...749...91F}, such as fallback \citep[e.g.][]{2021ApJ...920L..17V} and the pair-instability mass gap \citep[e.g.,][]{2002ApJ...567..532H,2017ApJ...836..244W}, as well as stellar winds \citep[e.g.][for a review]{2014ARA&A..52..487S}.

Finally, the predicted mass and velocity distribution of isolated BH lenses depends on the assumptions about the supernova.
We explored using an alternative natal kick prescription to gauge the effect on the isolated BH population (Appendix \ref{app:kicks}), which resulted in effectively identical yields but slower CoM speed ($\lesssim 50\ \rm{km\ s^{-1}}$).
However, supernova modeling in population synthesis can include variations on the remnant mass prescription or the natal kick distribution, which sometimes can be coupled or dependent on the evolutionary history of the system.
Such an exploration is beyond the scope of this work, but would be a natural extension in the exploration of stellar-mass BH lenses.

\subsection{Systematic errors}
Recently, between the submission of this manuscript and its acceptance, \cite{2022ApJ...937L..24M} analyzed the data as used in \cite{2022ApJ...933L..23L} and \cite{2022ApJ...933...83S} in order to investigate  the source of mass discrepancy in \event.
They infer a lens mass of $7.88\pm 0.82\ \rm{M_{\odot}}$ (see, e.g., Figure \ref{fig:mass_and_speed}), which is in agreement with \cite{2022ApJ...933...83S}.
They find that using new photometric OGLE reductions reduce the tension between astrometric and photometric measurements. 
Moreover, they find that the main source of uncertainty comes from blending from a close bright neighbor.
Finally, they suggest ``there is no strong evidence for systematic errors in the HST data reductions by \cite{2022ApJ...933...83S}" and claim that ``\cite{2022ApJ...933L..23L} reductions are affected by systematic errors".

\subsection{Open questions}
In our population, we consider all binaries to begin their evolution at rest with respect to their environment, and therefore any velocity changes are with respect to this zero-velocity inertial frame of reference.
We also assume the dynamical effects associated  with supernovae are the only way of disturbing that equilibrium. 
While massive stars and binaries are believed to be young ($<100$ Myr) and live their lives \textit{in situ}, disrupted NSs, BHs, and double compact objects can be as old as the Milky Way ($\approx 10$ Gyr).
We neglect the origin and long-term evolution of systems in the Galactic potential \citep[e.g.][]{2010ApJ...725L..91K}.
While the velocity dispersion of stars in the (thin) disk increases with time, stars are rarely heated above $50 \gtrsim \rm km\ s^{-1}$ in any direction \citep[e.g.][]{2007MNRAS.380.1348S}.
We do not account for systems formed in dense dynamical environments \cite[but see][]{2022ApJ...928..181K}.
However, MS-BH binaries have been detected in globular clusters \citep{2018MNRAS.475L..15G}.

\cite{2022ApJ...933...83S} associated \event\ to a young object, closer to us than to the Galactic disk; however, they do consider the possibility that is an older object that is just passing by the location of detection.
If \event\ is a young object formed in the field, particularly in the thin disk, our assumptions are nonetheless reasonable.
    
\subsection{Future prospects}
\label{subsec:microlensing}
Conventional gravitational microlensing methods lead to a degeneracy between the mass and the velocity of the lens.
Astrometric microlensing lifts this degeneracy by measuring the microlens parallax \citep[e.g.][]{2000ApJ...542..785G}, which led to the more precise mass estimates for \event.
\event\ has a low transverse speed ($\lesssim 45\ \rm{km\ s^{-1}}$) and is in the direction of the Galactic bulge \citep{2022ApJ...933...83S,2022ApJ...933L..23L}.
A lens with a low-magnitude natal kick is more likely to remain within the vicinities of its origin rather than emigrate to the outskirts of the Galaxy; this seems to be the case for \event.
Finally, we highlight that microlensing events could be comprised of binary lenses \citep{1991ApJ...374L..37M} or binary sources \citep{2017AJ....153..129J}, both of which lead to particular type of caustics.
These lens-source populations should be investigated in order to determine the most plausible configurations and what can we learn about the observed population. 

Although the field is far from being mature, sufficient progress has been made in identifying the most prominent binary configurations leading to microlensing of BHs and NSs. We have thus here  endeavored to outline some of the stellar avenues  that we believe
to be most relevant to interpreting microlensing events arising from BHs and NSs.

\begin{acknowledgments}
We thank Alexey Bobrick, Dan D'Orazio, Stephen Justham, Johan Samsing and Tom Wagg for useful discussions. 
We thank the Heising-Simons Foundation and the NSF (AST-1911206, AST-1852393, and AST-1615881) for support.
A.V-G. received support through Villum Fonden grant no. 29466.
\end{acknowledgments}

%

\vspace{5mm}


\software{ Scripts used for this study are publicly available in GitHub via \href{https://github.com/avigna/isolated-black-holes}{avigna/isolated-black-holes}.
          }



\appendix

\section{COMPAS setup}
\label{app:compas_setup}
Our rapid population synthesis simulation is done with COMPAS v02.27.05 \citep{2022ApJS..258...34R}.
The details of the initial conditions and setup are presented in Table \ref{tab:COMPAS}.

\begin{table*}
\caption{Initial values and default settings of the population synthesis simulation with COMPAS.
}
\label{tab:COMPAS}
\centering
\resizebox{\textwidth}{!}{%
\begin{tabular}{lll}
\hline  \hline
Description and name                                 														& Value/range                       & Note / setting   \\ \hline  \hline
\multicolumn{3}{c}{Initial conditions}                                                                      \\ \hline
Initial mass \monei                               															& $[5, 150]$\Msun    & \citet{2001MNRAS.322..231K} IMF  $\propto  {\monei}^{-\alpha}$  with $\alpha_{\rm{IMF}} = 2.3$ for stars above $5$\Msun	  \\
Initial mass ratio $\qi = \mtwoi / \monei $           												& $[0.01, 1]$                          &       Flat mass ratio distribution  $p(\qi) \propto  1$ with \mtwoi $\geq 0.1\Msun$   \\
Initial semi-major axis \ai                                            											& $[0.01, 1000]$\AU & Flat-in-log distribution $p(\ai) \propto 1 / {\ai}$ \\   
Initial metallicity \Zi                                           											& $[0.0001, 0.03]$                 & Representative metallicities (Section \ref{sec:method})        \\
Initial orbital eccentricity \ei                                 							 				& 0                                & Binaries circular at birth  \\
%
\hline
\multicolumn{3}{c}{Fiducial parameter settings:}                                                            \\ \hline
Stellar winds  for hydrogen rich stars                                   																&      \citet{2010ApJ...714.1217B}    &   Based on {\citet{2000A&A...362..295V,2001A&A...369..574V}}, including  LBV wind mass loss with $f_{\rm{LBV}} = 1.5$   \\
Stellar winds for hydrogen-poor helium stars &  \citet{2010ApJ...715L.138B} & Based on   {\citet{1998A&A...335.1003H}} and  {\citealt{2005A&A...442..587V}}  \\

%
Max transfer stability criteria & $\zeta$-prescription & Based on \citet[][]{2018MNRAS.481.4009V} and references therein     \\ 
Mass transfer accretion rate & thermal timescale & For stars  \citet[][]{2018MNRAS.481.4009V,2020MNRAS.498.4705V} \\ 
 & Eddington-limited  & For compact objects  \\
Non-conservative mass loss & isotropic re-emission &  {\citet[][]{1975MmSAI..46..217M,1991PhR...203....1B,1997A&A...327..620S}} \\ 
& &  {\citet{2006csxs.book..623T}} \\
Case BB mass transfer stability                                														& always stable         &       Based on  \citet{2015MNRAS.451.2123T,2017ApJ...846..170T,2018MNRAS.481.4009V}         \\ 
%
%
CE prescription & $\alpha-\lambda$ & Based on  \citet{1984ApJ...277..355W,1990ApJ...358..189D}  \\
CE efficiency $\alpha$-parameter                     												& 1.0                               &              \\
CE $\lambda$-parameter                               													& $\lambda_{\rm{Nanjing}}$                             &        Based on \citet{2010ApJ...716..114X,2010ApJ...722.1985X} and \citet{2012ApJ...759...52D}       \\
Hertzsprung gap (HG) donor in {CE}                       														& pessimistic                       &  Defined in \citet{2012ApJ...759...52D}:  HG donors don't survive a {CE}  phase        \\
%
%
{SN} natal kick magnitude \vk                          									& $[0, \infty)$\kms & Drawn from a Maxwellian distribution with an user-defined standard deviation ($\sigma_{\rm{rms}}^{\rm{1D}}$)          \\
 {SN} natal kick polar angle $\thetak$          											& $[0, \pi]$                        & $p(\thetak) = \sin(\thetak)/2$ \\
 {SN} natal kick azimuthal angle $\phi_k$                           					  	& $[0, 2\pi]$                        & Uniform $p(\phi) = 1/ (2 \pi)$   \\
 {SN} mean anomaly of the orbit                    											&     $[0, 2\pi]$                             & Uniformly distributed  \\
Core-collapse  {SN} remnant mass prescription          									     &  delayed                     &  From \citet{2012ApJ...749...91F}, which  has no lower {BH} mass gap  \\%
USSN  remnant mass prescription          									     &  delayed                     &  From \citet{2012ApJ...749...91F}   \\%
ECSN  remnant mass presciption                        												&                                 $m_{\rm{f}} = 1.26\Msun$ & Baryonic to gravitational mass relation using the equation-of-state from \citet{1996ApJ...457..834T}          \\
Core-collapse  {SN}  velocity dispersion $\sigma_{\rm{rms}}^{\rm{1D}}$ 			& 265\kms           & 1D rms value based on              \citet{2005MNRAS.360..974H}                          \\
 USSN  and ECSN  velocity dispersion $\sigma_{\rm{rms}}^{\rm{1D}}$ 							 	& 30\kms             &            1D rms value based on e.g.,    \citet{2002ApJ...573..283P,2004ApJ...612.1044P}    \\
PISN / PPISN remnant mass prescription               											& \citet{2019ApJ...882...36M}                    &       As implemented in \citet{2019ApJ...882..121S}      \\
Maximum NS mass                                      & $\rm{max}_{\rm{NS}} = 2.5$\Msun &  Following \citet{2012ApJ...749...91F}           \\
Tides and rotation & & Not included \\
\hline
\multicolumn{3}{c}{Simulation settings}                                                                     \\ \hline
Total number of binaries sampled per metallicity  & $5\times 10^6$                    &           \\
Sampling method                                      & Monte Carlo &      \\
Binary fraction                                      & $f_{\rm{bin}} = 1$ &    \\
Binary population synthesis code                                      & COMPAS &       \citet{2022ApJS..258...34R} \\
\hline \hline
\end{tabular}%
}
\end{table*}

\section{Alternative natal-kick distribution}
\label{app:kicks}

An understanding of the physics of supernovae is an ongoing endeavor \cite[see, e.g.,][for a recent review]{2021Natur.589...29B}.
In the context of binary population synthesis, supernova physics is one of the major uncertainties and is often parameterized in order to explore their impact in rates and distributions of, e.g., electromagnetic transients and gravitational-wave sources.
In rapid population synthesis, the details of supernovae cannot be incorporated, and therefore supernovae are simplified to instantaneous events where the mass, type, and natal kick of the remnant are determined by the structure of the progenitor.
For example, the supernova prescription used in the COMPAS default model \citep[][see also Appendix \ref{app:compas_setup}]{2012ApJ...749...91F} determines the compact-object remnant given the pre-supernova mass of the carbon-oxygen core and envelope (if any) of the stellar progenitor.
\cite{2012ApJ...749...91F} accounts for mass fallback to the newly born compact object, which results in two effects: increasing the mass and damping the natal kick of the compact-object remnant.
For BH formation, the natal kick distribution is damped with respect to the \cite{2005MNRAS.360..974H} velocity dispersion (Appendix \ref{app:compas_setup}); for the most massive ($\gtrsim 10\ \rm{M_{\odot}}$) stellar-mass BHs, complete fallback results in no natal kick.

We  explore the effect of an alternative natal kick distribution in our population.
We follow the natal kick formula from \cite{2016MNRAS.461.3747B}, where the kick velocity $v_{\rm{kick}}= \alpha (M_{\rm ejecta} / M_{\rm remnant}) + \beta$, with $\alpha=100$ and $\beta=-170$ \citep{2018MNRAS.480.5657B}, scales linearly with the ejecta mass.
This formula was originally introduced and studied in the context of young NS populations \citep{2016MNRAS.461.3747B}, but has since been used in the context of isolated black holes and microlensing events \citep[e.g.,][]{2019ApJ...885....1W,2020ApJ...905..134W}.
We choose this as an alternative model for the clear physical meaning, the simplicity of the implementation, and because the natal kick is decoupled from the remnant mass prescription, which allows us to compare exclusively the effect of the natal kick the resulting population.

\begin{figure}
	\includegraphics[trim=1cm 7.5cm 1cm 7.5cm,clip,width=0.5\columnwidth]{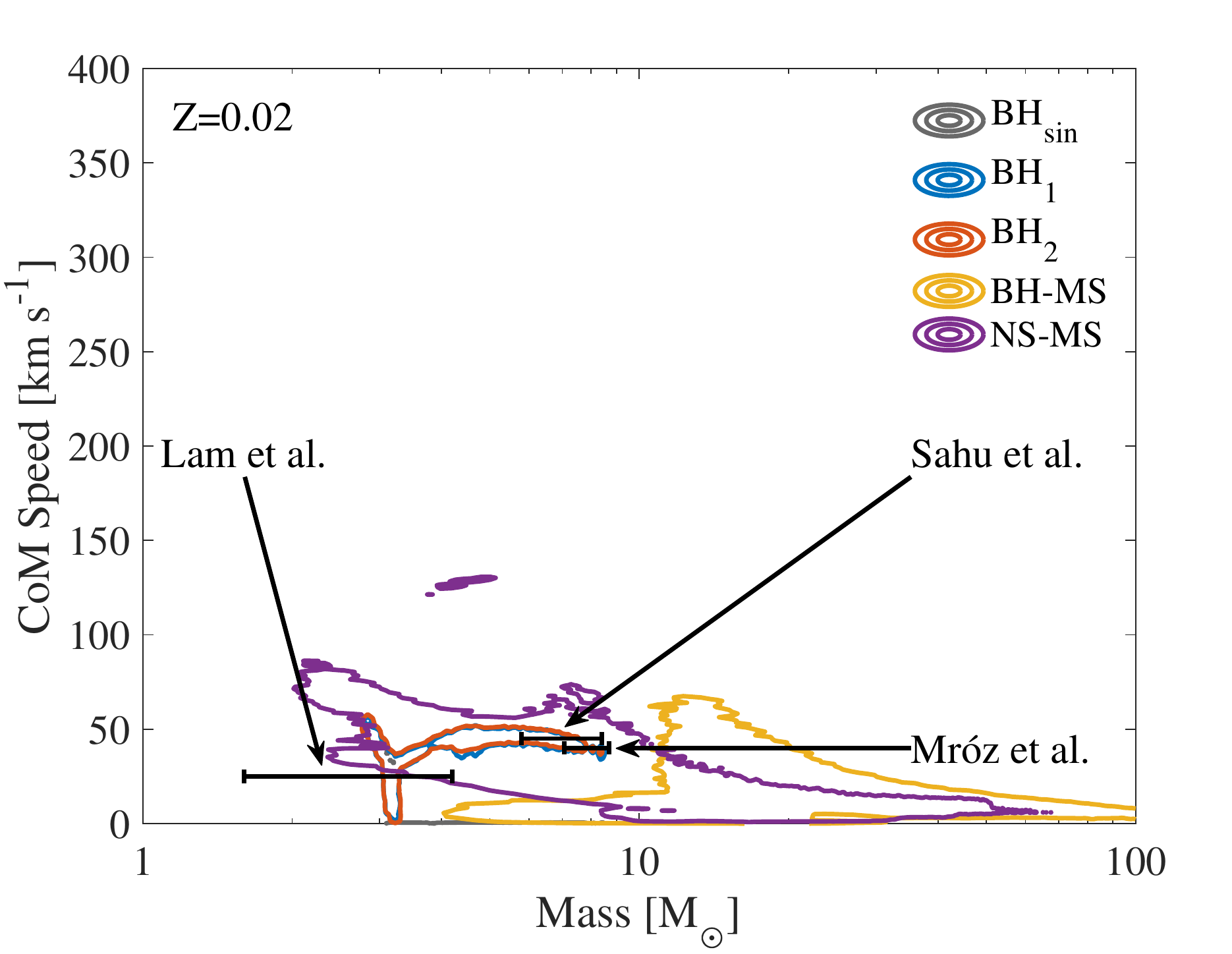}
    \includegraphics[trim=1cm 7.5cm 1cm 7.5cm,clip,width=0.5\columnwidth]{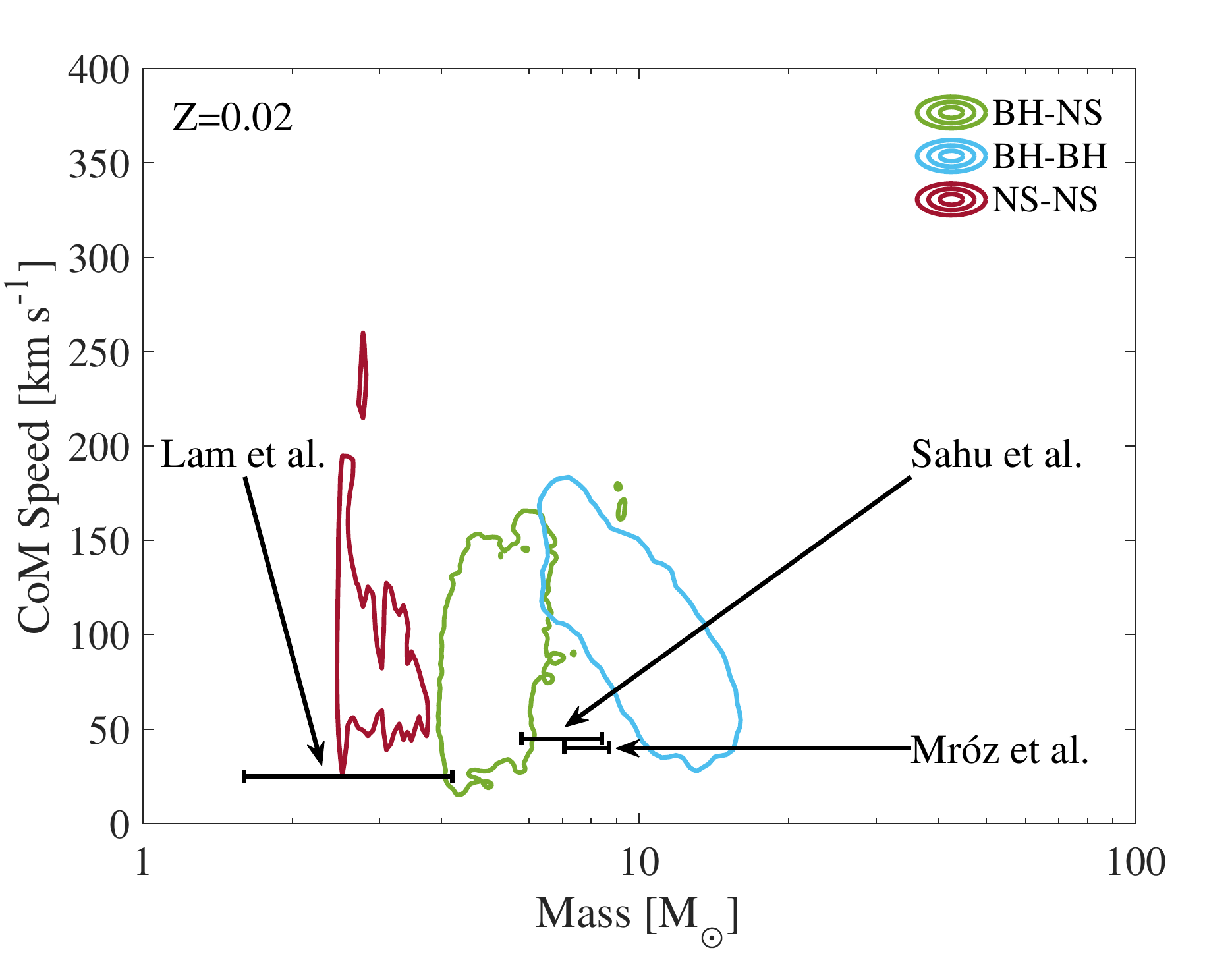}
    \caption{
    Results using the (alternative) natal kick model from \cite{2016MNRAS.461.3747B}.
    Besides the natal kick distribution, all other assumptions are identical to the default model (Section \ref{app:compas_setup}).
    Additionally, the layout here is identical to that of Figure \ref{fig:mass_and_speed}, which allows for a direct comparison (more details on Appendix \ref{app:kicks}).
    }
    \label{fig:app:kicks}
\end{figure}

We create an identical population to our default model, at $Z=0.02$, with the only difference that we obtain the natal kick from the aforementioned model \citep{2016MNRAS.461.3747B,2018MNRAS.480.5657B} instead of sampling it from a (default) Maxwellian distribution \citep{2005MNRAS.360..974H}.
First, we explore how the yields of the sub-populations of interest are affected by this alternative natal kick distribution.
For the isolated BH population, the yields are effectively identical to those from our default model.
For the BH-MS and NS-MS population, the yields increase at most by a factor of a few with respect to our default model.
However, for the double compact object populations, the yields from the alternative model can be up to an order of magnitude larger than for our default model; however, these rates are still at least an order of magnitude less than those from isolated BHs.
Figure \ref{fig:app:kicks} shows the sub-populations of interest in the mass vs CoM speed parameter space (cf. Figure \ref{fig:mass_and_speed}).
The effect of the alternative natal kick distribution is mostly noticeable in the isolated BH population.
While the mass range of isolated BHs remains similar with respect to the default model, the CoM is decreased drastically ($\lesssim 50\ \rm{km\ s^{-1}}$); however, the sub-populations of isolated BHs ($\rm{BH_{sin}}$, $\rm{BH_{1}}$, and $\rm{BH_{2}}$) are still consistent with one another, occupying the same parameter space.
This is a consequence of more massive BHs having less ejecta, resulting in low-kicks \citep[for complete fallback, the formula from][can result in negative kick magnitudes, which we cap to 0]{2016MNRAS.461.3747B,2018MNRAS.480.5657B}.
This model favors a low-speed isolated BH population.
For the BH-MS and NS-MS population, the alternative natal kick model widens both the mass and CoM speed distributions, but only slightly.
Finally, the mass and CoM speed distributions of double compact objects are similar in both natal kick models.

\bibliography{Manuscript}{}
\bibliographystyle{aasjournal}



\end{document}